\documentclass[structabstract]{aa}

\usepackage{txfonts}
\usepackage{graphicx}
\usepackage{natbib}
\usepackage{aalongtable}
\usepackage{url}
\bibpunct[; ]{(}{)}{,}{a}{}{;} 

\begin{document}

\title{Spectroscopic and photometric oscillatory envelope variability during the S\,Doradus outburst of the Luminous Blue Variable R71\thanks{Based on observations collected at ESO's Very Large Telescope under Prog-IDs: 69.D-0390(D), 289.D-5040(A), 290.D-5032(A), 091.D-0116(A,B), 092.D-0024(A), 094.D-0266(A,B,C), 096.D-0043(A,B,C), 097.D-0006(A,B), 598.D-0005(A,B) and at the MPG/ESO 2.2-m Telescope under Prog-IDs: 076.D-0609(A), 078.D-0790(B), 086.D-0997(A,B), 087.D-0946(A), 089.D-0975(A), 094.A-9029(D), 096.A-9039(A), 097.D-0612(A,B), 098.D-0071(A).}
\fnmsep 
} 


\author{
A. Mehner\inst{1}
\and D. Baade\inst{2} 
\and J.H. Groh\inst{3}
\and T. Rivinius\inst{1}
\and F.-J. Hambsch\inst{4,5}
\and E.S. Bartlett\inst{1}
\and D. Asmus\inst{1}
\and C. Agliozzo\inst{6,7} 
\and T. Szeifert\inst{1} 
\and O. Stahl\inst{8} 
} 

\institute{ESO -- European Organisation for Astronomical Research in the Southern Hemisphere, Alonso de Cordova 3107, Vitacura, Santiago de Chile, Chile 
  \and ESO -- European Organisation for Astronomical Research in the Southern Hemisphere, Karl-Schwarzschild-Stra{\ss}e 2,  85748 Garching, Germany
   \and School of Physics, Trinity College Dublin, Dublin 2, Ireland
   \and AAVSO -- American Association of Variable Star Observers, Cambridge, USA
   \and ROAD -- Remote Observatory Atacama Desert, Vereniging Voor Sterrenkunde (VVS), Oude Bleken 12, B--2400 Mol, Belgium   
   \and Millennium Institute of Astrophysics (MAS), Nuncio Monseñor S\'{o}tero Sanz 100, Providencia, Santiago, Chile   
   \and Departamento de Ciencias Fisicas, Universidad Andres Bello, Avda. Republica 252, Santiago, 8320000, Chile
   \and Zentrum f\"ur Astronomie der Universit\"at Heidelberg, Landessternwarte, K\"onigstuhl 12, 69117 Heidelberg, Germany
}   


\abstract {Luminous Blue Variables (LBVs) are evolved massive stars that exhibit instabilities not yet understood. Stars can lose several solar masses during this evolutionary phase. The LBV phenomenon is thus critical in our understanding of the evolution of the most massive stars.} {The LBV R71 in the Large Magellanic Cloud is presently undergoing an S\,Doradus outburst, which started in 2005. To better understand the LBV phenomenon, we determine R71's fundamental stellar parameters during its quiescence phase. In addition, we analyze multi-wavelength spectra and photometry obtained during the current outburst.} {Pre-outburst CASPEC spectra from 1984--1997, EMMI spectra in 2000, UVES spectra in 2002, and FEROS spectra from 2005 are analyzed with the radiative transfer code CMFGEN to determine the star's fundamental stellar parameters. A spectroscopic monitoring program with VLT X-shooter since 2012 secures visual to near-infrared spectra throughout the current outburst, which is well-covered by ASAS and AAVSO photometry. Mid-infrared images and radio data were also obtained.} {During quiescence, R71 has an effective temperature of $T_{\mathrm{eff}} = 15\,500$~K and a luminosity of log$(L_*/L_{\odot})= 5.78$. We determine its mass-loss rate to $4.0 \times 10^{-6}~M_{\odot}$~yr$^{-1}$.  We present R71's spectral energy distribution from the near-ultraviolet to the mid-infrared during its present outburst. Semi-regular oscillatory variability in the star's light curve is observed during the current outburst. Absorption lines develop a second blue component on a timescale twice that length. The variability may consist of one (quasi-)periodic component with $P \sim 425/850$~d with additional variations superimposed.} {R71 is a classical LBV, but at the lower luminosity end of this group. Mid-infrared observations suggest that we are witnessing dust formation and grain evolution. During its current S\,Doradus outburst, R71 occupies a region in the HR diagram at the high-luminosity extension of the Cepheid instability strip and exhibits similar irregular variations as RV\,Tau variables. LBVs do not pass the Cepheid instability strip because of core evolution, but they develop comparable cool, low-mass, extended atmospheres in which convective instabilities may occur. As in the case of RV\,Tau variables, the occurrence of double absorption lines with an apparent regular cycle may be due to shocks within the atmosphere and period doubling may explain the factor of two in the lengths of the photometric and spectroscopic cycles.} 


\maketitle

\section{Introduction}
\label{intro}

LBVs, also known as S\,Doradus variables, are evolved massive stars that exhibit instabilities which are not yet understood (\citealt{1984IAUS..105..233C,1997ASPC..120..387C,1994PASP..106.1025H,1997ASPC..120.....N}, and references therein). They represent a brief but critical phase in massive star evolution, because several solar masses can be expelled during this stage (see, e.g. \citealt{2014A&A...564A..30G}).   The LBV phenomenon extends to luminosities as low as log$(L / L_{\odot})\sim5.2$, corresponding to stars with initial masses of $\sim$20~$M_{\odot}$. LBVs experience outbursts with enhanced mass loss during which they appear to make transitions in the Hertzsprung-Russell (HR) diagram from their quiescent hot state ($T_{\textnormal{\scriptsize{eff}}}\sim16\,000$--$30\,000$~K) to lower temperatures ($T_{\textnormal{\scriptsize{eff}}}\sim8\,000$~K).   
Their instability could be responsible for the empirically found upper luminosity boundary above and to the right of which no supergiants are found \citep{1979ApJ...232..409H,1984Sci...223..243H}. The lowest luminosity LBVs may have had a Red Supergiant (RSG) phase prior to becoming LBVs.

LBVs have been generally considered to be stars in transition to the Wolf-Rayet stage (e.g., \citealt{1983A&A...120..113M,1994PASP..106.1025H,1994A&A...290..819L}). Recent observational and theoretical work suggests that some LBVs can be the immediate progenitors of supernovae (SNe). Some core-collapse SNe may be boosted to very high luminosity when the ejecta shock circumstellar matter expelled during a previous LBV outburst (\citealt{2007ApJ...656..372G,2009Natur.458..865G,2007ApJ...666.1116S,2008ApJ...686..467S}; see \citealt{2012MNRAS.423L..92Q} for an alternative explanation).
\citet{2013A&A...550L...7G} showed using stellar evolutionary models that single rotating stars with initial mass in the range of 20--25$~M_{\odot}$ have spectra similar to LBVs before exploding as SNe. This result sets the theoretical ground for low-luminosity LBVs to be the endpoints of stellar evolution. 
\citet{2007AJ....133.1034S} discussed the possibility that the progenitor of SN~1987A, the Blue Supergiant Sk~-69~202, was in an LBV phase when exploding. However, Sk~-69~202 had a lower luminosity, log$(L / L_{\odot})\sim 5.01$ \citep{1989ApJ...341..925B}, than any other confirmed LBV.\footnote{Alternative progenitor hypotheses for SN~1987A include a  merger model \citep{1992PASP..104..717P} or an interacting winds model \citep{2008A&A...488L..37C}.}

Most of the fundamental questions about the physical cause of the LBV instability are still unsolved. 
Outbursts with visual magnitude variations of 1--2~mag and constant bolometric luminosity are commonly referred to as classical LBV outbursts. During giant eruptions, such as P\,Cygni in the 17th century (e.g., \citealt{1988IrAJ...18..163D,1992A&A...257..153L}) and $\eta$~Car in the 1840s (e.g., \citealt{1997ARA&A..35....1D,2012ASSL..384.....D}), the visual magnitude increases by more than 2~mag and the bolometric luminosity likely increases. 
The most promising explanations for the LBV instability mechanism involve radiation pressure instabilities, but also turbulent pressure instabilities, vibrations and dynamical instabilities, and binarity cannot be ruled out (\citealt{1994PASP..106.1025H} and references therein). Especially the impact of binarity has received much attention lately (e.g., \citealt{2014ApJ...796..121J,2015MNRAS.447..598S,2016A&A...593A..90B}).
It is also being debated if the observed radius change of the photosphere can be a pulsation or outer envelope inflation driven by the Fe opacity bump \citep{2012A&A...538A..40G,2015A&A...580A..20S}.  

The Luminous Blue Variable (LBV) R71 (= HD 269006) in the Large Magellanic Cloud (LMC) is presently undergoing an S\,Doradus outburst that started in 2005 \citep{2009IAUC.9082....1G}. In 2012, the star had reached unprecedented visual brightness, accompanied by remarkable variations in its optical spectrum \citep{2012CBET.3192....1G,2013A&A...555A.116M}. R71's outburst is well-sampled with multi-wavelength spectroscopic and photometric observations and  is excellent for a case study toward a better understanding of the LBV phenomenon. The star had several LBV outbursts in the last century. Between 1900 and 1950 the star experienced two contiguous outbursts \citep{2012IAUS..285...29G,2017AJ....154...15W}. Another outburst was observed between 1970--1977 \citep{1974MNRAS.168..221T,1975A&A....41..471W,1979A&AS...38..151V,1981A&A...103...94W,1982A&A...112...61V}. Currently, R71 reached $V$ magnitudes of more than one magnitude brighter than during these previous outbursts and the event was thus termed ``supermaximum'' by \citet{2017AJ....154...15W}.\footnote{\citet{2017AJ....154...15W} regret the observational gap from the early 1990s to 2000. CASPEC spectra cover this period well (see Section \ref{obs:spec}). The star is at its minimum phase during this period.}
Because of the scarce observational coverage of LBV outbursts and their potential importance in massive star evolution this large-amplitude event calls for special attention.

\begin{figure*}
\centering
\resizebox{\hsize}{!}{\includegraphics{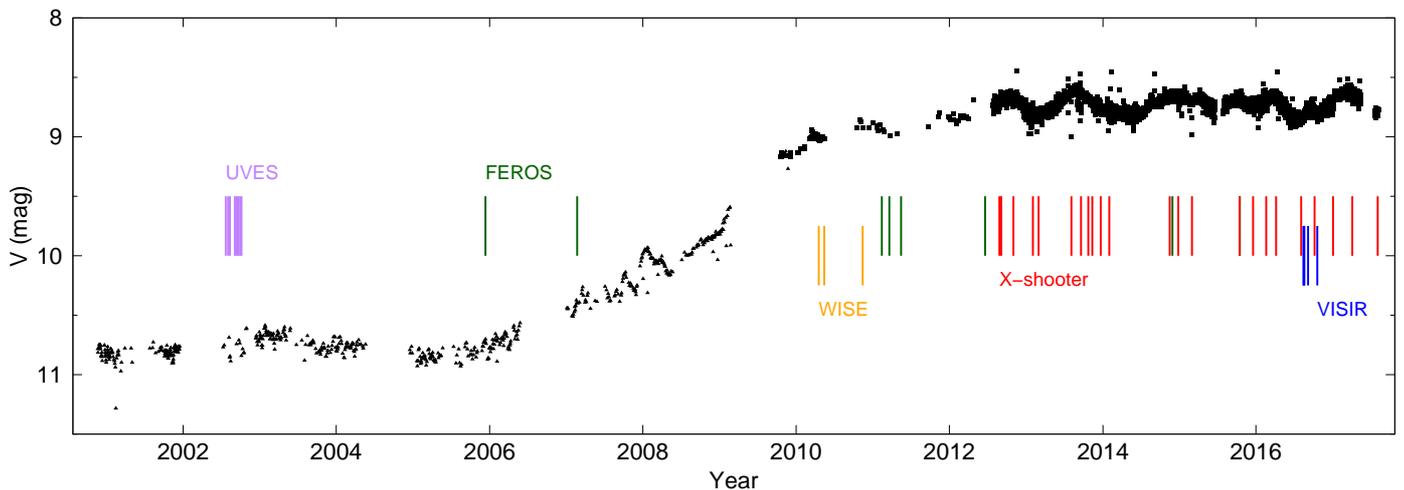}}
     \caption{R71's visual light curve using data from ASAS (triangles; \citealt{1997AcA....47..467P}) and from AAVSO (squares; \protect\url{www.aavso.org}). Epochs at which spectroscopic observations with UVES, FEROS, and X-shooter are available are indicated. The epochs of WISE and VISIR mid-infrared imaging observations are also marked. An oscillation in the light curve is clearly apparent since 2010. Variations with a shorter timescale can be seen in 2007 and 2008 during the onset of the outburst.}
     \label{figure:lightcurve}
\end{figure*}

The first years of R71's current outburst were reviewed in \citet{2013A&A...555A.116M}. R71's visual light curve reached its maximum in 2012 with $m_{\textnormal{\scriptsize{V,2012}}} = 8.7$~mag -- a brightening of about 2~mag compared to its quiescent state (Figure \ref{figure:lightcurve}). With small variations, the star has maintained this brightness since and shows no symptoms of an imminent end of the outburst.  Its spectrum resembles a late-F supergiant with an unusually low apparent temperature for an LBV outburst of only $T_{\textnormal{\scriptsize{eff,2012}}} \sim$~6\,650~K. The rise in R71's visual magnitude and the low effective temperature are unprecedented for this star. 

In this second paper we cover aspects of R71's current outburst not discussed in the first paper, such as spectral and photometric oscillations and the evolution of its ultra-violet to mid-infrared spectral energy distribution. We also settle the debate on its luminosity \citep{1993SSRv...66..207L,1981A&A...103...94W}, which has an impact on its evolutionary path \citep{1994PASP..106.1025H}. Classical LBVs have $M_{\textnormal{\scriptsize{bol}}} < -9.6$~mag and have very likely not been RSGs. Less luminous LBVs have $M_{\textnormal{\scriptsize{bol}}} = -8$~mag to $-9$~mag, lower temperatures, smaller amplitudes of their outbursts, and lower mass loss rates. There is no strong evidence of a separation between less luminous and classical LBVs, but see \citet{2004ApJ...615..475S} for a discussion of a potential deficiency of LBVs in the luminosity range between them. 

In Section \ref{obs} we describe the observations and the details of the CMFGEN radiative transfer modeling and the search for periodicities in the light curve. In Section \ref{results} we present R71's stellar parameters during quiescence, its spectral energy distribution, and the oscillations observed in both the light curve and the spectrum during the current outburst. In Section \ref{discussion} we discuss its evolutionary state and the oscillatory variations. In Section \ref{conclusion} we summarize our conclusions.

\section{Observations and data analysis}
\label{obs}

\subsection{Spectroscopic data}
\label{obs:spec}

In 2012 we started a spectroscopic monitoring campaign of R71's S\,Doradus outburst with X-shooter at the Very Large Telescope (VLT). X-shooter is a medium-resolution echelle spectrograph that simultaneously observes with three arms the wavelength region from $3\,000$--$24\,800$~\AA\ \citep{2011A&A...536A.105V}. 
Spectra are obtained with the narrowest available slits of 0\farcs5 in the UVB arm, 0\farcs4 in the VIS arm, and 0\farcs4 in the NIR arm yielding spectral resolving powers of $R\sim9\,000-17\,000$. In addition, spectra with 5\arcsec\ slits in all three arms provide us with the means to achieve relative flux calibration, absolute flux calibrations to 10\% accuracy, and to investigate the spectral energy distribution. A journal of the observations is presented in Table \ref{table:journal}.

Table \ref{table:journal} also lists additional archival spectroscopic observations of R71 during the years 1984-2015. 
VLT Ultraviolet and Visual Echelle Spectrograph (UVES; \citealt{2000SPIE.4008..534D}) observations were obtained in 2002 during R71's quiescent state to investigate microvariations over two and a half months. The spectra cover the wavelength region from $3\,280$--$6\,690$~\AA\ with spectral resolving power of $R\sim40\,000$.  MPG ESO-2.2m Fiber-fed Extended Range Optical Spectrograph (FEROS; \citealt{1999Msngr..95....8K}) observations in 2005--2015 were obtained throughout the current outburst starting at its onset. FEROS is fed by two fibers with a 2\arcsec\ aperture for simultaneous spectra of the target and sky. The wavelength range is $3\,500$--$9\,200$~\AA\ and the spectral resolving power is $R\sim48\,000$. Each spectroscopic data set was reduced with the corresponding ESO pipeline (X-shooter pipeline version 2.8.4, UVES pipeline version 5.7.0, FEROS pipeline version 1.0.0). The spectra were corrected for the instrumental response and atmospheric extinction but not for interstellar extinction and telluric lines.
La Silla 3.6m Cassegrain ESO Echelle Spectrograph (CASPEC; \citealt{1979Msngr..17...27L}) data from 1984--1997 cover the wavelength region from $3\,740$--$5\,445$~\AA\ with spectral resolving power of $R\sim20\,000$. One La Silla 3.6m ESO Multi-Mode Instrument (EMMI; \citealt{
1986SPIE..627..339D}) spectrum from 2000 covers the wavelength region from $6\,240$--$6\,870$~\AA\ with spectral resolving power of $R\sim20\,000$.  International Ultraviolet Explorer (IUE; \citealt{1976MmSAI..47..431M}) spectra around 2\,800~\AA\ obtained during minimum phase in 1981, 1985, and 1986 are used to confirm R71's terminal wind velocity.
Enhanced data products of the Spitzer InfraRed Spectrograph (IRS; \citealt{2004ApJS..154...18H}) observations between March 2004 and March 2005 were obtained from the NASA/IPAC Infrared Science Archive. The mean full width at half maximum of the extraction profile ranges between 3\arcsec\ and 8\arcsec.

In \citet{2013A&A...555A.116M}, we used the [\ion{Fe}{II}] emission lines in the 2002 UVES spectra to determine R71's systemic velocity to v$_{\textnormal{\scriptsize{sys}}} =$~192$\pm$3~km~s$^{-1}$. All radial velocities stated in this paper are relative to the systemic velocity.

\subsection{Photometric data}
\label{obs:phot}

The available photometry during R71's current outburst is exceptional. No other LBV outburst has been recorded with such a high temporal cadence and duration.
Figure \ref{figure:lightcurve} shows R71's visual light curve since 2000 indicating the epochs of spectroscopic observations and recent mid-infrared imaging observations. The $V$-band photometry was retrieved from the All Sky Automated Survey (ASAS; \citealt{1997AcA....47..467P}) and from the American Association of Variable Star Observers (AAVSO; www.aavso.org).

The $V$-band photometry from AAVSO was acquired since July 2012 with a frequency of 1--2 days at the Remote Observatory Atacama Desert (ROAD; \citealt{2012JAVSO..40.1003H}) with an Orion Optics, UK Optimized Dall Kirkham 406/6.8 telescope and a FLI 16803 CCD camera and Astrodon Photometric $V$ and $I_C$ filters.
Twilight sky-flat images were used for flat-field corrections. The reduction was performed with the MAXIM DL program (\url{http://www.cyanogen.com}) and magnitudes were determined with the LesvePhotometry program (\url{http://www.dppobservatory.net}). The magnitude error is about 0.02~mag.

Figure \ref{figure:SED_all} shows R71's spectral energy distribution from the near-ultraviolet to the infrared. Photometry during the star's quiescent phase was retrieved from many catalogs using VIZIER \citep{2000A&AS..143...23O}. In the mid-infrared, we complement this data set by observations during the current outburst obtained with the upgraded VLT spectrometer and imager for the mid-infrared (VISIR; \citealt{2004Msngr.117...12L,2015Msngr.159...15K,2016SPIE.9908E..0DK}) and with The Wide-field Infrared Survey Explorer (WISE). WISE observed R71 between April and November 2010 and the magnitudes were retrieved from the AllWISE Data Release \citep{2013yCat.2328....0C}.

\begin{table}
\caption{R71's mid-infrared flux densities obtained from VISIR observations between 14 August and 19 October 2016.\label{table:mid-infrared}}
\begin{tabular}{cccc}
\hline\hline
filter & wavelength & flux density & error flux density \\ 
	&  ($\mu$m) & (mJy) & (mJy) \\
	\hline
M-BAND$^a$	&	4.82		&	430	&	39	 \\
PAH1		&	8.59	 &	159	&	26	 \\
SIV$^b$	&	10.49	&	245	&	20 \\
SIV\_2	&	10.77	&	687	&	135 \\
PAH2	&	11.25	&	363	&	54	 \\
PAH2\_2	&	11.88	&	375	&	32 \\
NeII\_1	&	12.27	&	612	&	146 \\
NeII	$^c$	&	12.81	&	293	&	46 \\
Q1	&	17.65	&	2948		&	338 \\
Q2	&	18.72	&	3625		&	600	 \\
\hline
\multicolumn{4}{l}{$^a$ The individual measurements and conversion factor are all} \\
\multicolumn{4}{l}{\phantom{$^b$}  consistent with each other; we exclude a calibration problem.} \\
\multicolumn{4}{l}{$^b$ The source detection is very faint; only aperture photometry can} \\
\multicolumn{4}{l}{\phantom{$^b$} be performed.} \\
\multicolumn{4}{l}{$^c$ The source detection is very faint; the flux is uncertain.} \\
\end{tabular}
\end{table}

\begin{table*}
  \begin{center}
    \caption{Summary of ATCA radio observations and radio map properties.$^a$}
    \begin{tabular}{lccccccc}
      \hline
      \hline
      date & frequency & integration time & synthesized beam  & position angle & largest angular scale & rms noise & $F_{\nu}^b$ \\
      \small
      & (GHz) & (min) & HPBW (arcsec) & $(\rm deg)$ & (arcsec) &$(\rm \mu Jy\, beam^{-1})$ & $\rm (\mu Jy)$\\
      \hline
      2014 Feb 14-16& 5.5 & 135 &2.33$\times$1.76 & -12.2 & 100 & 13 & $<39$\\
      2014 Feb 14-16& 9.0  & 135 &1.35$\times$1.17 & -18.5 & 50& 26 & $<78$\\
      2015 Mar 2-4& 5.5 & 180 &2.05$\times$1.79 & -3.6 & 50& 8 & $<24$\\
      2015 Mar 2-4& 9.0 & 180 &1.49$\times$1.32  & -12.8 &30 & 9 & $<27$\\
\hline
\multicolumn{8}{l}{$^a$ Project ID: C1973} \\
\multicolumn{8}{l}{$^b$ R71 is not detected. Upper limits on the flux density at the radio frequencies. } \\
    \end{tabular}
    \label{tab:ATCA}
    \\
  \end{center}
\end{table*}

R71 was observed with VISIR in standard imaging mode with perpendicular nodding several times between 14 August and 19 October 2016. 
Three epochs in the PAH2\_2 (11.9\,$\mu$m) filter were obtained on
2016-08-14 with 30~min on source exposure time, on 2016-08-18 with 53~min, and on 2016-09-05 with 21~min. Another epoch in the Q1 filter (17.7\,$\mu$m) on 2016-08-17 yields 10\,min on source exposure. Finally, on 2016-10-19, the star was observed in the filters: M-BAND
(4.8\,$\mu$m; 20\,min), PAH1 (8.6\,$\mu$m; 5\,min), SIV (10.5\,$\mu$m; 2\,min), SIV\_2
(10.8\,$\mu$m; 2\,min), PAH2 (11.3\,$\mu$m; 2\,min), NEII (12.8\,$\mu$m; 2\,min), NEII\_1
(12.3\,$\mu$m; 2\,min), and Q2 (18.7\,$\mu$m; 2\,min).
All observations were carried out at an airmass of 1.5 or higher, which negatively affected the image quality.
For flux calibration, the science observations were either preceded or followed by a mid-infrared standard star  \citep{1999AJ....117.1864C}, which is the main source of uncertainty in the resulting photometry.
The data reduction was performed with a custom made \textsc{python} pipeline and flux
measurements were obtained using the custom developed \textsc{idl} software package
\textsc{mirphot} \citep{2014MNRAS.439.1648A}.
The photometry was determined using both classical aperture photometry with an aperture of $1\arcsec$ and Gaussian fitting to the point source.
The difference between these two methods gives an estimate of the measurement uncertainty. The systematic flux calibration uncertainty is about 10\% if only one flux standard was obtained. The resulting mid-infrared spectral flux densities from the 2016 VISIR images of R71 are listed in Table \ref{table:mid-infrared}.

\subsection{Radio data}

R71 was observed at radio wavelengths with the Australia Telescope Compact Array at two different epochs in 2014 and 2015, see Table \ref{tab:ATCA}. The observations were performed with the array in the most extended configuration (6~km) and the Compact Array Broadband Backend (CABB) ``4~cm'' receiver. The 
receiver bandwidth was split in two 2-GHz sub-bands centered at
\hbox{5.5\,GHz} and \hbox{9\,GHz}, with 2048$\times$1-MHz channels in each sub-band. R71 was observed with 15~min scans on target, alternating with 2.5~min on the phase calibrator ICRF~J052930.0$-724528$. The scans were distributed over 12 values of hour angle in order to achieve a good \emph{uv} coverage. The total integration times are listed in Table \ref{tab:ATCA}. For bandpass and flux calibrations, observations of ICRF~J193925.0$-634245$ were performed on each date. 

The data were reduced separately for each central frequency (5.5 and 9 GHz) with the MIRIAD package \citep{1995ASPC...77..433S}. The data reduction process consisted in flagging bad data and calculating bandpass corrections and complex gain solutions.
The calibrated visibilities were imported in the Common Astronomy Software Applications (CASA) package v5.0.0 \citep{2007ASPC..376..127M} and imaged with the task \texttt{tclean}. A natural weighting scheme for the visibilities was adopted in order to preserve the sensitivity. For the deconvolution of the dirty image, the algorithm of \citet{1974A&AS...15..417H} was used. Table \ref{tab:ATCA} summarizes the properties of the cleaned radio maps. The rms noise in the residual maps is less than a factor of two larger than the theoretical noise, meaning that the sensitivity is mostly limited by statistical noise. R71 is not detected. Upper limits on the flux density at the radio frequencies are provided in the last column of Table \ref{tab:ATCA}. 

\subsection{CMFGEN radiative transfer modeling}
\label{obs:CMFGEN}

The stellar winds of LBVs can be so dense that many emission lines form in the wind and veil the underlying photospheric spectrum. The radiation emitted by the stellar surface interacts with the optically-thick, dense wind, making the analysis of the emerging stellar spectrum challenging. 
In addition to temperature, luminosity, and effective gravity, other quantities such as mass-loss rate, microturbulence, and wind acceleration law affect the morphology of the spectrum. Complex radiative transfer models, which include the necessary physics to study the radiation transport across the atmosphere and wind, are needed to obtain realistic parameters.

We analyzed R71's pre-outburst spectra with atmosphere models computed with the radiative transfer code CMFGEN (version 5may17, \citealt{1998ApJ...496..407H}). CMFGEN comprises the state-of-the-art in non-local thermodynamic equilibrium (non-LTE) radiative transfer and has been successfully applied to several LBVs (e.g., \citealt{2001ApJ...553..837H,2006ApJ...638L..33G,2009ApJ...698.1698G,2011ApJ...736...46G,2009A&A...507.1555C,2012A&A...541A.146C,2011AJ....142..191G}). In CMFGEN, each model is defined by the effective temperature, luminosity, mass-loss rate, wind volume filling factor, wind terminal velocity, effective gravity (or stellar mass), and by the abundances.
For a review on the essential properties of the code and the basics of spectroscopic analysis of massive stars using photospheric and wind diagnostics, see, e.g., \citet{2011JPhCS.328a2020G} and \citet{2011BSRSL..80...29M}.
The following diagnostics were used to find the best-fitting stellar model for R71:

\begin{itemize}

\item {\it Effective Temperature.}
The effective temperature is obtained using the ionization balance of different ionization stages of the same chemical species. The best line diagnostic for R71 and our spectral coverage is the Si ionization balance using the lines \ion{Si}{II} $\lambda$6347; \ion{Si}{III} $\lambda\lambda$4553, 4568, 4575; \ion{Si}{IV} $\lambda\lambda$4089, 4116. 
We also use the \ion{Mg}{II} $\lambda\lambda$4481, 6343 lines. For both elements, we assume LMC abundance. LMC abundances are scaled to the solar abundances from \citet{2009ARA&A..47..481A}, assuming  $Z_{\odot}=0.014$ and $Z_{LMC}=0.006$ \citep{2013MNRAS.433.1114Y}.

\item {\it Luminosity and extinction.}
The bolometric luminosity and extinction are obtained self-consistently by fitting CMFGEN model atmosphere spectra to the observed optical to near-infrared spectral energy distribution during the star's quiescence phase. 
The distance to the LMC is well-determined and the errors associated with this method depend largely on the quality of the photometry. We adopted a distance to the LMC of $49.97 \pm 0.19~ \textnormal{(statistical)} \pm 1.11~ \textnormal{(systematic)}$~kpc \citep{2013Natur.495...76P}. We iteratively matched the colors and fluxes of the $JHK$ photometry to those synthesized from reddened CMFGEN models. Because the extinction in the near-infrared is low, we avoid large uncertainties in the luminosity, which may arise from an uncertain ratio of extinction to reddening $R_V$. To uncover a potential unusual extinction curve toward R71, we also determine the ratio of extinction to reddening (Section \ref{results:extinction}). 

\item {\it Gravity and mass.}
The wings of the hydrogen Balmer and Paschen lines are the main diagnostics for the gravity, log$~g$.

\item {\it Mass-loss rate and clumping.}
Wind clumping has a significant impact on obtaining the mass-loss rates of massive stars \citep{1991A&A...247..455H,2006ApJ...637.1025F,2008A&ARv..16..209P,2013MNRAS.428.1837S,2014A&A...568A..59S}. CMFGEN allows for the presence of clumping using a volume-filling approach, assuming that material is unclumped close to the star and reaches maximum clumpiness at large distances.
In principle, the mass-loss rate and clumping could be constrained from the emission strengths and electron scattering wings of recombination lines such as H$\alpha$ and higher Balmer lines. However, we lack precise response curves for the pre-outburst spectroscopic data and thus cannot constrain the clumping. We therefore use a volume-filling factor of $f = 1$ ($\bar{\rho} = f \rho$, where $\bar{\rho}$ is the homogeneous unclumped wind density). The mass-loss rate scales with $\dot{M} = const. \sqrt{f}$. Accordingly, the derived mass-loss rates are upper limits.  

\item {\it Surface abundance.}
It is essential to determine the abundance of He and the CNO elements to constrain R71's history and evolutionary state. LBVs show usually an overabundance of He and N, and a depletion of H, C, and O. The He abundance is determined from the non-diffuse \ion{He}{I} lines at $\lambda\lambda$3965, 4121, and 4713~\AA, but also taking into account other helium lines. C and N abundances are derived using the numerous C and N lines present in the optical range, e.g., \ion{N}{II} $\lambda\lambda$3995, 4601--4643, 5667--5711,6482, \ion{N}{III} $\lambda\lambda$4196, 4216, 4511, 4515, 4523, 4602, 4907, and \ion{C}{II}  $\lambda\lambda$4267, 6578, 6583, \ion{C}{III} $\lambda\lambda$4070, 4153--56--63, 4326, 5305, 6205. 
The O abundance is estimated using \ion{O}{I} $\lambda\lambda$7772--7775.
We only determine the abundances of H, He, and CNO. For all other elements we assume LMC abundances. As above for the effective temperature, solar abundances are from \citet{2009ARA&A..47..481A} and the LMC abundances are scaled to the solar abundances, assuming  $Z_{\odot}=0.014$ and $Z_{LMC}=0.006$.

\item{\it Microturbulence.}
Line broadening reflects the effects of both rotation and turbulence. 
Microturbulence, $v_{turb}$, can have a considerable effect on the abundances. We constrain it using the \ion{Si}{III} lines at $\lambda$4560~\AA. A large effect can be observed for the \ion{He}{I} $\lambda\lambda$6678, 7065 lines, where, with lower microturbulence, the red line wings move bluewards.

\item{\it Wind terminal velocity.}
We determine a lower limit to the terminal velocity, $v_{\infty}$, in optical spectra using the velocity of the maximum absorption of hydrogen and helium lines. 
We validate this estimation using the metallic resonance line \ion{Mg}{II} 2803 in IUE spectra obtained in 1981, 1985, and 1986 during quiescent state.

\end{itemize}

\subsection{Search for periodicities in the light curve}

The AAVSO $V$-band light curve during R71's outburst (since 2012) was searched for periodicities using the Lomb-Scargle technique \citep{1976Ap&SS..39..447L,1982ApJ...263..835S}. The Lomb-Scargle periodogram is a commonly used statistical tool designed to detect periodic signals in unevenly spaced observations. A heuristic was used to determine the frequency range and the grid spacing, resulting in a period range of $\sim$0.2--20\,000~d and a grid spacing of 0.00010893~d$^{-1}$. The analysis was performed using the \textsf{astropy.stats.LombScargle} package.

We generated 1000 simulated light curves via a bootstrap technique. Bootstrapping involves sampling with replacement from the original dataset. Magnitude-time data pairs are selected at random from the original light curve with multiple selections possible. Multiple selections of the same pairs are then discarded, leaving us with light curves made up of ~63\% of the original data. Lomb-Scargle analysis was performed on each of these light curves in the same manner as on the original light curve.

\section{Results}
\label{results}

\subsection{R71's current LBV outburst}
\label{results:eruption}

We analyzed changes in R71's near-ultraviolet to mid-infrared brightness during its current LBV outburst. Several spectral features between 1984 and 2017 are also compared. The spectra until 2005 cover R71's quiescence phase, see Figure \ref{figure:lightcurve}. Small variations in the line profiles occur during quiescence phase due to microvariations, also observed in the photometry, but the spectra closely resemble each other. At the time of the 2005 FEROS spectrum, the current outburst commenced but yet without major changes in the spectrum. 
The 2007 FEROS spectrum is a snapshot of the transitional phase between the quiescent and the eruptive state. The FEROS spectra since 2011 and the X-shooter spectra since 2012 show R71 in outburst, when the visual magnitude had reached its maximum visual brightness.

\subsubsection{The spectral energy distribution}
\label{results:SED}

\begin{figure*}
\centering
{\includegraphics[width=0.95\textwidth]{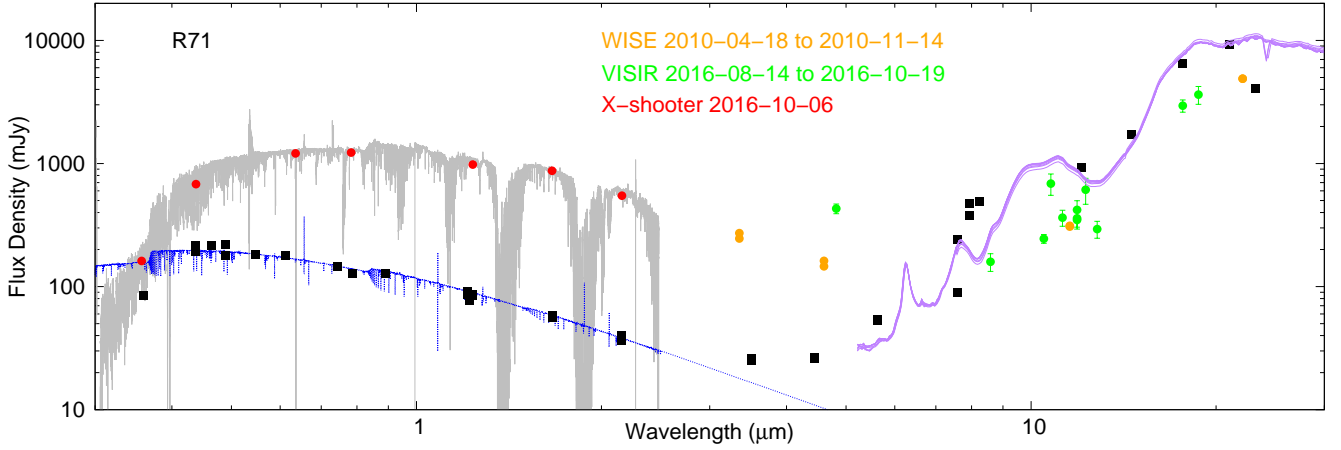}}
     \caption{R71's near-ultraviolet to mid-infrared SED. Literature values of R71's quiescent state are retrieved from VIZIER (black squares). The best-fit, reddened CMFGEN model for R71's quiescent state is displayed (blue curve), as well as the October 2016 X-shooter spectrum during outburst (gray curve) and the synthetic $U$- to $K$-band photometry (red circles). 2016 VISIR $M$- to $Q2$-band photometry (green circles) and 2010 WISE photometry (orange circles) complement the SED during R71's current LBV outburst to the mid-infrared. Comparison to Spitzer/IRS spectra in 2004--2005 suggests that the dust in the outermost regions has cooled down in the last decade (purple curve).}
     \label{figure:SED_all}
\end{figure*}
\begin{figure*}
\centering
{\includegraphics[width=0.29\textwidth]{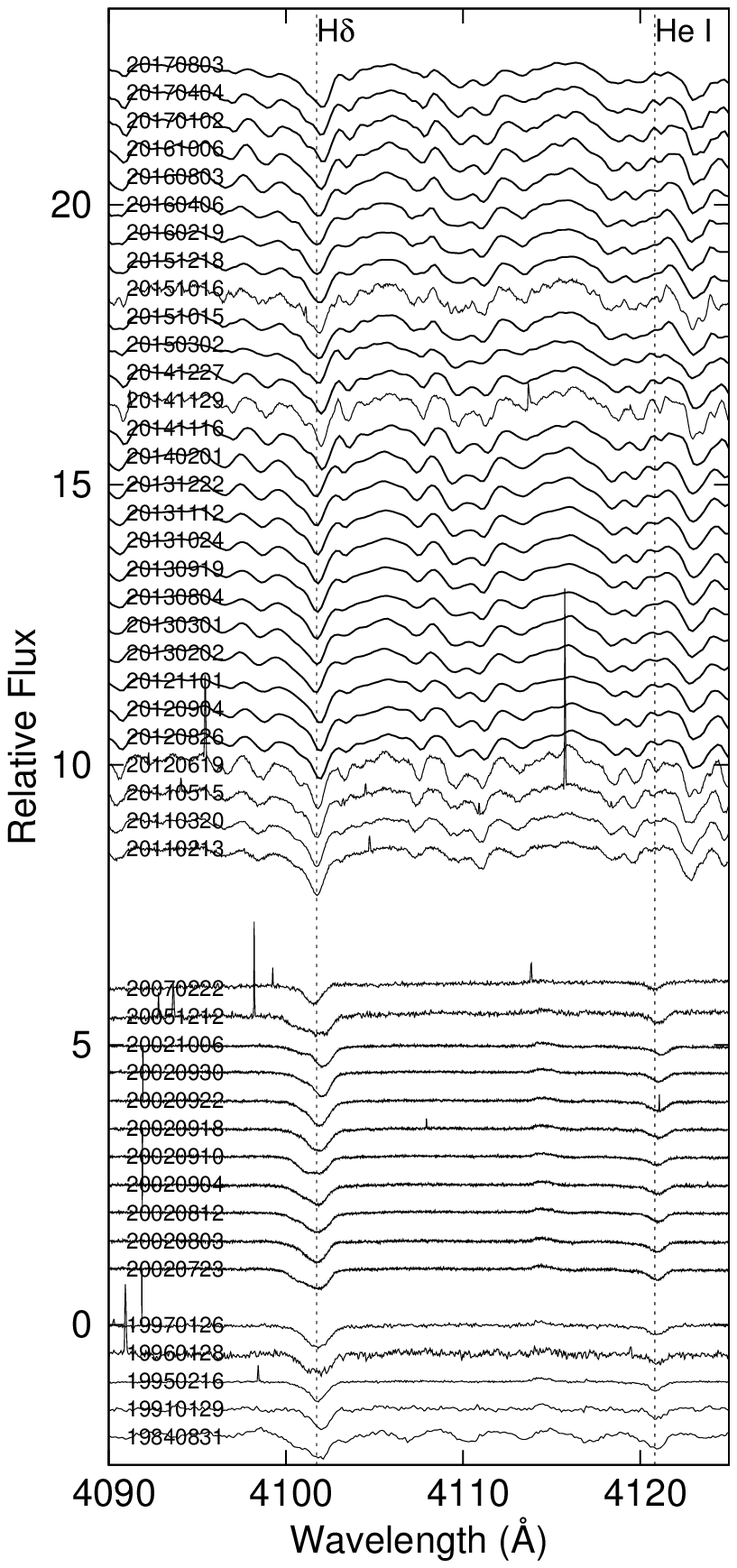}\includegraphics[width=0.29\textwidth]{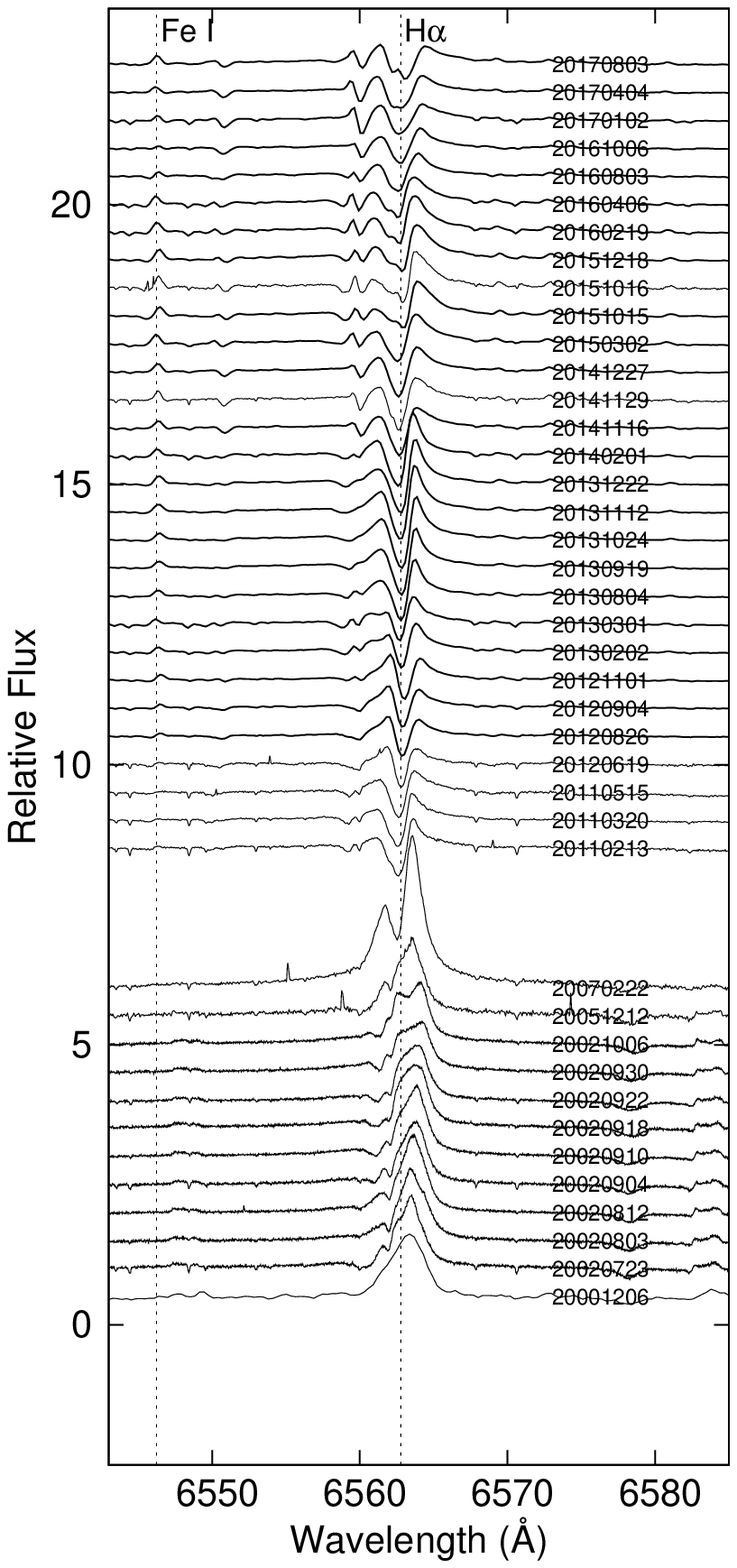}
\includegraphics[width=0.29\textwidth]{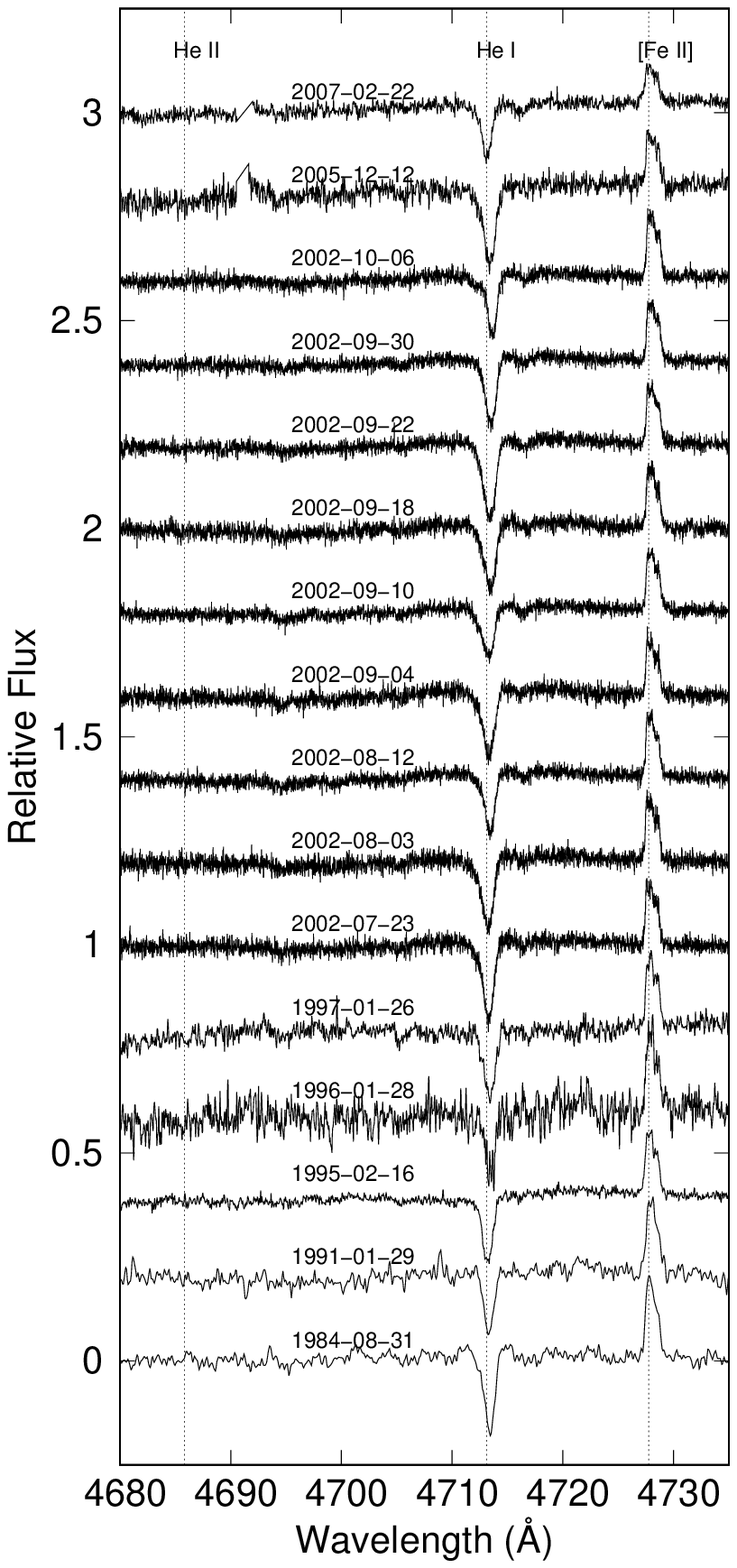}}
     \caption{{\it Left and middle panel (common vertical timescale):} H$\delta$ and H$\alpha$ in R71 from 1984 to 2017. The H$\alpha$ P\,Cyg profile changed to an inverse double-peaked symmetric profile during the outburst. Until 2007, the star is hot and increased line emission implies an increased amount of circumstellar material. An absorption component close to system velocity dominates the profile during the outburst phase and several absorption components trace the line-of-sight structure.  An example of a low-excitation emission \ion{Fe}{I} line is shown. {\it Right panel:} \ion{He}{II} 4686 and \ion{He}{I} 4713 in R71 from 1984 to 2007. No \ion{He}{II} emission is present. (The continuum is normalized to unity, velocities are in R71's rest frame.)}
\label{figure:Hdelta_Halpha_HeII}
\end{figure*}

Figure \ref{figure:SED_all} compares R71's pre-outburst near-ultraviolet to mid-infrared spectral energy distribution (SED) with its SED during the current S\,Doradus outburst. 
During the outburst, R71's spectral energy density increases dramatically from about 3\,500~\AA\ to 6--7~$\mu$m. Bluewards of 3\,500~\AA, the star becomes fainter. This is the usual behavior seen in S~Doradus outbursts, i.e., the flux at ultraviolet wavelengths is reprocessed to longer wavelengths as the effective temperature of the photosphere decreases. The bolometric luminosity likely remains constant.

Redward of about 7~$\mu$m, where the mid-infrared excess from surrounding cold dust dominates, lower flux is observed in 2010 and 2016 compared to 2004--2005. This suggest cooling of the dust in the outermost regions. The VISIR and WISE observations suggest that the emission bump around 10~$\mu$m has disappeared. This indicates that grain evolution is occurring and that we are possibly seeing grain growth (e.g., \citealt{2006ApJ...639..275K}).

In $M$ band ($\lambda$4.82~$\mu$m) we find an increase of hot dust emission between the WISE observations in 2010 and the VISIR images in 2016. Since the individual measurements and conversion factor for the VISIR observations are consistent with each other, we exclude a calibration problem. The increase in $M$ band may thus be the result of newly formed dust caused by the current outburst. In this case, we expect an increase at longer wavelengths in the next years. 
Two of the VISIR photometric points (SIV\_2, NEII\_1) exhibit significantly higher
fluxes than other simultaneously recorded filters at nearby wavelengths. As the same
calibrator stars are used and their cross-calibration is consistent, calibration issues can be excluded. Future low-resolution VISIR spectra would resolve the issue.

With respect to the oscillatory variations discussed in Section \ref{results:cycle} we find that during the brighter phases in $V$ the color $V-I$ is smaller, i.e., the star becomes bluer. Synthetic photometry of the X-shooter spectra with wide 5\arcsec\ slits confirms this for all wavelengths.  The $V-I$ color varies by about $\pm0.065$~mag from the average value $V-I \approx 0.558$~mag. This results in temperature variations of about 500~K between the (local) minima and maxima (changes in spectral type between F5 and F9; \citealt{2013ApJS..208....9P}). The associated radius variation is less than 10\%. 

 \begin{figure*}
\centering
\includegraphics[width=0.935\textwidth]{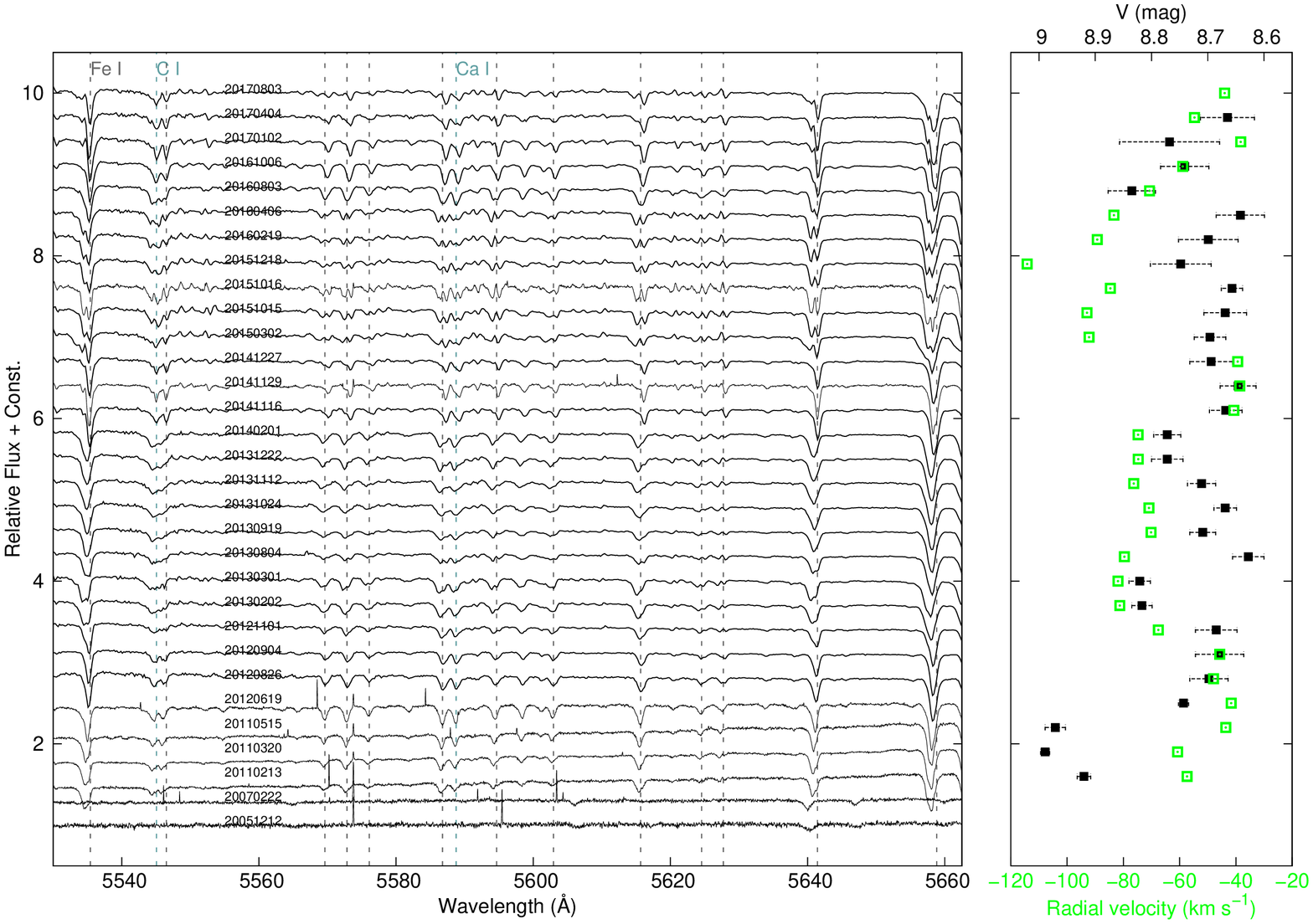}
     \caption{Left panel: Absorption lines develop a second blue component with a cycle of about 850~d. Three cycles have been observed so far, see also Figure \ref{figure:overview_time}. Gray dotted lines indicate \ion{Fe}{I} lines. Right panel: $V$-band magnitude (black squares) at the time of the spectra to the left (average of $\pm1$~d around the time of the spectrum). The radial velocity of the blue wing with respect to the red wing of the line is shown (green squares, measurements were performed at the mid-point between maximum and minimum depths of the lines).}
     \label{figure:overview}
\end{figure*}

\subsubsection{Hydrogen and helium lines} 
\label{results:hydrogen_helium}

Figure \ref{figure:Hdelta_Halpha_HeII} shows a time series of hydrogen and helium lines from 1984 to 2017.
H$\alpha$ changed from a P\,Cyg profile with broad emission wings  to a prominent double-peaked profile by 2011. 
In 2007 extra emission compared to the continuum is observed indicating an increased mass-loss rate. 
A strong blue-shifted absorption component at $v \approx -120$~km~s$^{-1}$ is present since November 2014. One could debate if this may have been caused by a shell ejection or a time-dependent wind, perhaps triggered by the bi-stability mechanism \citep{2011A&A...531L..10G}. Assuming that the extra H$\alpha$ emission observed in 2007 may have been the signature of a mass-ejection event, a corresponding shell of material may now be at about 250~au.
The helium absorption lines disappear as the star becomes cooler. There is no evidence of \ion{He}{II} emission during its quiescent state, which could have indicated a wind-wind interaction with a potential close companion.

\subsubsection{Oscillatory variations in the light curve and the spectrum}
\label{results:cycle}

During R71's current outburst, oscillatory variations are observed in its optical light curve and spectrum (Figures \ref{figure:lightcurve}, \ref{figure:overview}, and \ref{figure:overview_time}). These variations were also mentioned by \citealt{2017AJ....154...15W}, who determined a periodicity of $445\pm40$~d.
The light curve during outburst shows crudely sinusoidal variations between fainter and brighter states. In Section \ref{results:SED} we already described the $V-I$ color variations during these oscillations; from the color variations we infer temperature variations of about 500~K and radius variation of less than 10\%. In 2007 and 2008, when the star's brightness was increasing, the variations are observed but with shorter timescales.

 \begin{figure}
\centering
\resizebox{\hsize}{!}{\includegraphics{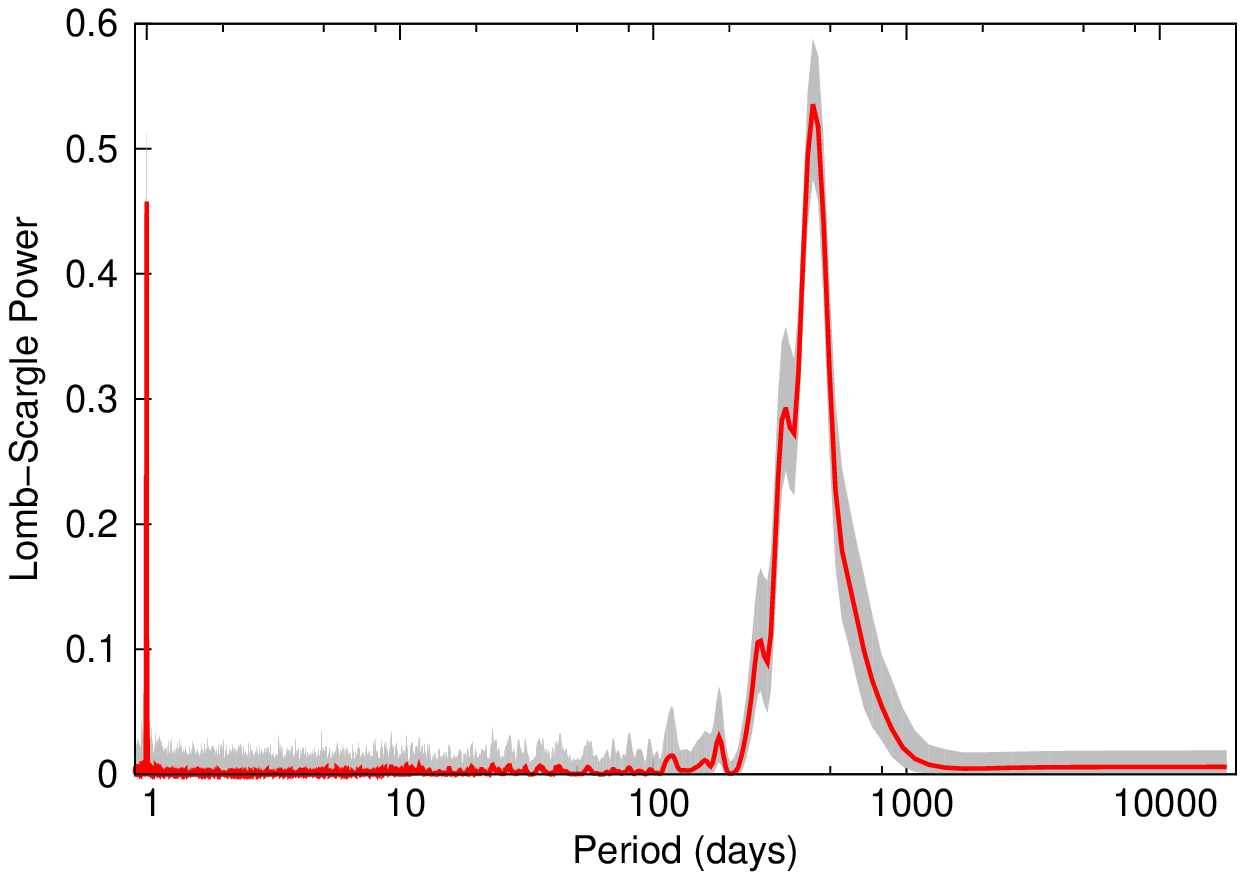}}
     \caption{Lomb-Scargle periodogram of R71's $V$-band light curve from 2012--2017. The grey shaded contours represent the range of Lomb-Scargle powers obtained with 1000 simulated light curves. A period of 425~d is found.}
     \label{figure:periodogram}
\end{figure}

 \begin{figure}
\centering
\resizebox{\hsize}{!}{\includegraphics{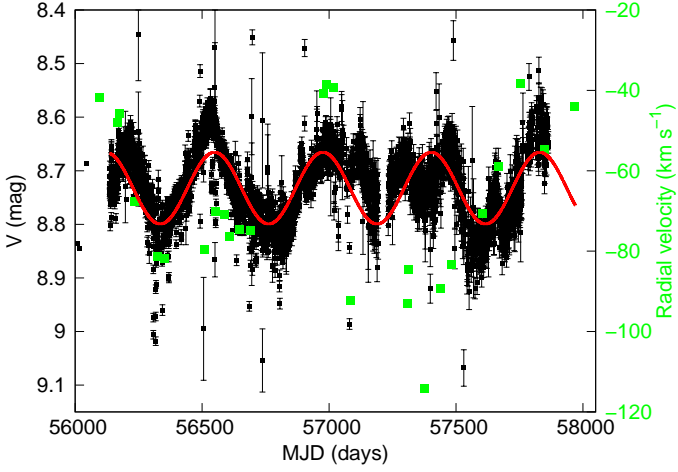}}
     \caption{R71's $V$-band magnitude (black squares) and radial velocities of the blue wing with respect to the red wing of the double line profiles (green squares) are shown. A sinusoidal curve with a period of 425~d, corresponding to the peak in the periodogram, is superposed in red. The photometric cycle is half the length of the spectroscopic cycle with some chaotic behavior. This may hint at period doubling, see Section \ref{discussion:pulsations}.}
     \label{figure:overview_time}
\end{figure}

 \begin{figure}
\centering
\resizebox{0.5\textwidth}{!}{\includegraphics{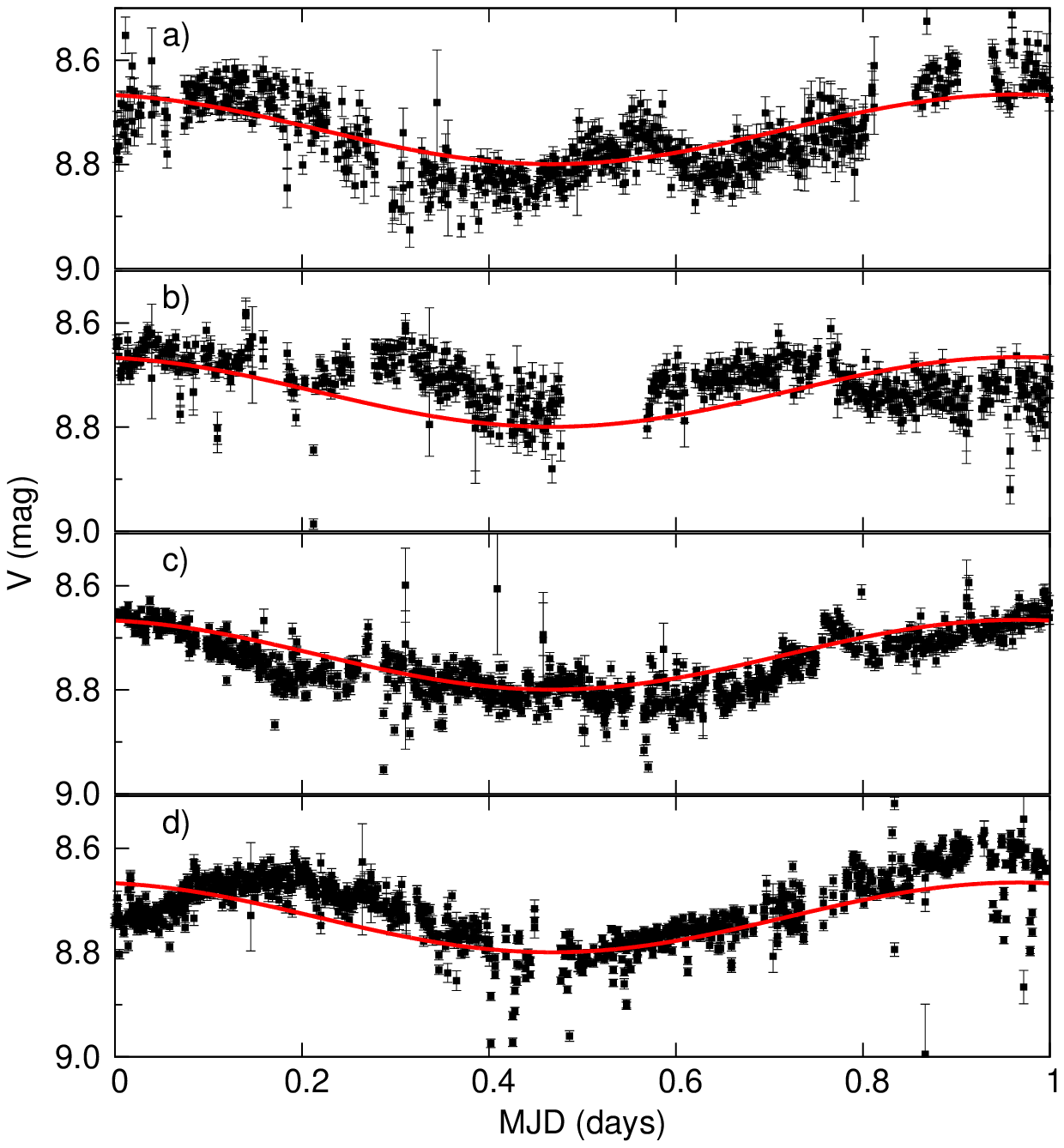}}
     \caption{R71's $V$-band magnitude (black squares) folded in phase space with a cycle of 425~d. A sinusoidal curve with a period of 425~d is shown in red. Cycle a) corresponds to MJD 57417--57844, cycle b) to MJD 56990--57417, cycle c) to MJD 56564--56990, and cycle d) to MJD 56135--56564.}
     \label{figure:overview_cycle}
\end{figure}

Figure \ref{figure:periodogram} shows the Lomb-Scargle periodogram of the star's $V$-band light curve from 2012--2017. The periodogram reveals strong peaks at frequencies corresponding to periods of about 0.2~d, 0.5~d, 1.0~d, and 425~d. The three shortest periods correspond to the duration of the observation, the nightly cadence, and half of the nightly cadence. Lomb-Scargle analysis was performed on 1000 bootstrapped light curves in the same manner as on the original light curve. In all instances, a period at 425~d was recovered, indicating that this value is robust and significant at a $\gtrsim$3-sigma level. 
While the light curve and this analysis indicate pseudo-steady oscillations, these are to some extent chaotic, which points toward a phenomenon within the extended atmosphere.
A quantitative estimate of the period error is not possible, and would be meaningless to some extent, as we cannot distinguish between a random error and a systematic evolution of the period during the time span of the observations.
 
Spectroscopic variations manifest themselves in that absorption lines vary between single and double on a timescale of about 850~d, twice the length of the photometric cycle. The uncertainty of the spectroscopic cycle length is about 50~d, limited by the sparsity of spectroscopic epochs. The spectroscopic variation can best be observed in the many \ion{Fe}{I} absorption lines (Figure \ref{figure:overview}).  The bluer of the two absorption lines is at a velocity of about $-70$~km~s$^{-1}$. The right panel of Figure \ref{figure:overview} and Figure \ref{figure:overview_time} correlate the radial velocity of the  blue wing of the \ion{Fe}{I} absorption with the star's $V$-band magnitude. Three spectroscopic cycles have been observed since 2011.  The spectra in 2011 have two absorption components, which become single by June 2012. In November 2012, the second absorption component reappears and then disappears again by November 2014, etc. The cycle length is twice as long as the photometric cycle, hinting at period doubling, see Section \ref{discussion:pulsations}.

\subsubsection{Low-excition emission nebula and dust shell}
\label{nebula}

In 2012, many low-excitation metal emission lines appeared and became stronger (\citealt{2013A&A...555A.116M}; see also the \ion{Fe}{I} line indicated in the middle panel of Figure \ref{figure:Hdelta_Halpha_HeII}). \citet{2013A&A...555A.116M} proposed that these lines indicate the presence of a  neutral nebula with an expansion velocity on the order of a few  km~s$^{-1}$, likely ejected during the current outburst. However, it now seems more likely that these emission lines are due to dynamical processes in the extended envelope. This is supported by the evolution observed in metal lines such as \ion{Fe}{II} from P\,Cyg profiles to inverse P\,Cyg profiles \citep{2017AJ....154...15W}, which has also been seen for the LBV S\,Doradus \citep{1990A&A...235..340W}.

Already before the current eruption, R71 showed nebular lines, such as [\ion{N}{II}] \citep{1986A&A...158..371S}, and strong mid-infrared radiation \citep{1984MNRAS.209..759G,1986A&A...164..435W,1999A&A...341L..67V}, which are evidence of a dust shell, but \citet{2003A&A...408..205W} found no extended optical nebula in the {\it HST/} WFPC2 image of R71 obtained in 1998. If R71 formed an optical LBV nebula during its 1970s outburst, it is either extremely faint or very small ($< 0.5$\arcsec).   \citet{2014A&A...569A..80G} discuss extensively R71's infrared spectral features, and derive dust shell parameters, composition, and mass. They modeled R71's optical to infrared spectral energy distribution and derived an outer radius of the dust shell of $6 \times 10^4$~au (1.2\arcsec\ on sky).
In case of a mass ejection event in 2007, the resulting shell would be at a distance of about 250~au from the star (5~mas on sky at the distance of the LMC) and thus would only be detectable with high-angular resolution techniques such as interferometry (see Figure 1 in \citealt{2016A&ARv..24....6B}).

The mid-infrared VISIR images suffer from bad image quality, mainly because of the high airmass, but also because of large seeing values during the observations, making the detection of a small nebula difficult. The best image quality was achieved in the PAH2\_2 image on 2016-08-18 and in Q1 on 2016-08-17. In both images the source has a featureless roundish extent with a FWHM of $\sim 1\arcsec$ (in agreement with the outer dust radius of 1.2\arcsec\ derived by \citealt{2014A&A...569A..80G}). Because of the uncertainty in image degradation with airmass due to a degraded chopping performance of the telescope, this value should be regarded as an upper limit to the true size of the object. No evidence of further extended emission is found in any of the images.

IRS spectra in 2004 and 2005 (Figure \ref{figure:SED_all}) show that the 8--12~$\mu$m wavelength range is dominated by a strong silicate emission feature, demonstrating the presence of an optically thin or clumpy dust shell heated by the star. A weaker corresponding silicate feature is present at $18~\mu$m and polycyclic aromatic hydrocarbon (PAH) emission is visible at 6.3$~\mu$m, 7.7$~\mu$m, and 11.3$~\mu$m. Evolution of these mid-infrared features indicates that we are witnessing grain evolution, see Section \ref{results:SED}.

LBV nebulae are often ionized and free-free emission is observed at radio wavelengths. Successful observations of extended LBV nebulae in the LMC with ATCA have been reported \citep{2012MNRAS.426..181A}. However, R71 is not detected in 2014--2015 ATCA radio maps. To explain this non-detection, we derive the radius $R_{s}$ of the Str{\"o}mgren sphere for ionized hydrogen in a nebula surrounding R71 during the quiescent phase (as the ionizing source). We adopt an ultraviolet photon flux of log$(N_{UV}) \sim 45.5$ (the value for a B3 supergiant; \citealt{1973AJ.....78..929P}) and assume a typical electron density of $100\,\rm cm^{-3}$ for an LBV nebula. This results in $R_{s}\approx0.2$~pc ($0.9\arcsec$ on sky). A similar HII region, in thermal equilibrium with electron temperature of $T_\mathrm{e}\approx 10^{3}-10^{4}$~K, produces flux densities on the order of 1~mJy at 9~GHz and would be detected with ATCA. We deduce that the ionized region in R71's ejecta must thus be much smaller than $<0.35\arcsec$ to escape detection with the achieved sensitivities. Note that for higher electron densities the ionized region would be even smaller, but brighter. For sufficiently high electron densities ($\approx 5\times10^{4}\,\rm cm^{-3}$), it is possible that hydrogen recombined since the beginning of the outburst.

\subsection{Extinction and reddening toward R71}
\label{results:extinction}

\begin{figure}
\centering
\resizebox{\hsize}{!}{\includegraphics{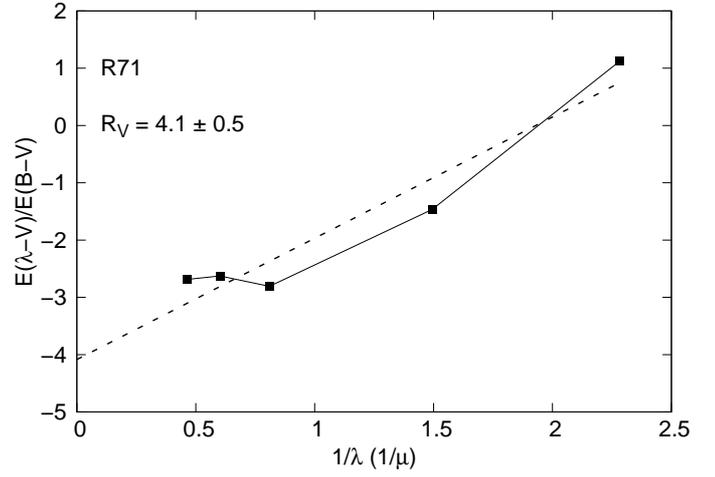}}
     \caption{Linear fit to the extinction curve of R71. $B$- to $R$-band photometry were taken from \citet{2002yCat.2237....0D} and $J$- to $K$-band photometry from \citet{2003yCat.2246....0C}. For the star's true magnitudes in these filters, we synthesized photometry from our best-fit unreddened CMFGEN model (gray curve in Figure \ref{figure:CMFGEN_SED}). The resulting ratio of extinction to reddening is $R_V = 4.1\pm0.5$.}
     \label{figure:RV}
\end{figure}

\begin{figure}
\centering
\resizebox{\hsize}{!}{\includegraphics{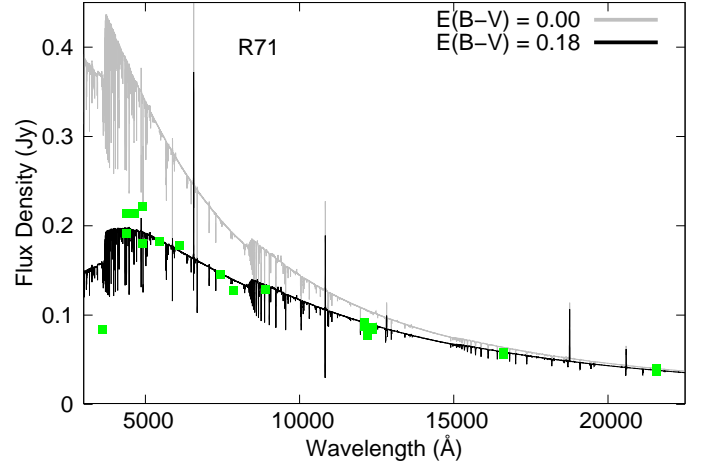}}
     \caption{Best-fit CMFGEN atmosphere model compared to R71's optical to near-infrared photometry. The best match with the photometry (green squares, retrieved from VIZIER) is found for $E(B-V)=0.18$~mag. The three outliers at wavelengths blueward of 5000~\AA\ are ignored since these correspond most likely to photometry obtained in phases where the star is not at its quiescent state.}
     \label{figure:CMFGEN_SED}
\end{figure}

The extinction and reddening toward an object depend on the wavelength, the amount of dust along the line of sight, and the dust grain sizes and composition. The value of the ratio of extinction to reddening $R_V = A_V/E(B-V)$ implies the size of the dust grains. The Galactic interstellar dust in the line of sight to the LMC does not significantly affect the spectrum of R71 \citep{1986AJ.....92.1068F} and its spectral classification can thus be used to measure the amount of reddening and extinction. 

We used $B$- to $R$-band photometry from \citet{2002yCat.2237....0D}, $J$- to $K$-band photometry from \citet{2003yCat.2246....0C}, and for the star's true magnitudes in these filters we synthesized photometry from our best-fit unreddened CMFGEN model. 
Figure \ref{figure:RV} shows the linear fit to the extinction curve resulting in $R_V = 4.1 \pm 0.5$, which is higher than the average of $R_V = 3.4$ for the LMC \citep{2003ApJ...594..279G}. However, the extinction curve suggests that there is no abnormal reddening. We lack precise simultaneous multi-wavelength photometry of the star's quiescent state to confidently determine the $R_V$ value and thus assume the average LMC value. However, a value of $R_V$ different from average may not be unexpected for a star that has undergone several S\,Doradus phases accompanied by high mass-loss \citep{1993SSRv...66..207L,1999A&A...341L..67V}. Unusual grain size and dust distribution have, e.g., also been observed for $\eta$~Car \citep{1986MNRAS.222..347M,1992A&A...262..153H} and AG\,Car \citep{2009ApJ...698.1698G}. 

The process to derive the reddening and extinction toward R71 is iterative, as mentioned in Section \ref{obs:CMFGEN}. 
Figure \ref{figure:CMFGEN_SED} shows that our best-fit CMFGEN model spectrum reddened by $E(B-V)=0.18$~mag  and $R_V = 3.4$ according to the Fitzpatrick law  \citep{1999PASP..111...63F} matches the observed photometry best. This results in an extinction value of $A_V = 0.6 \pm 0.1$~mag.
An extinction value of $A_V = 0.6$~mag is similar to that found by \citet{1993SSRv...66..207L}, who derived $A_V = 0.63$~mag, and much higher than the values found by \citet{1981A&A...103...94W}  and \citet{1988A&AS...74..453V}, who derived $A_V = 0.15$~mag and $A_V = 0.37$~mag, respectively. These authors assumed much smaller reddening values and thus found lower effective temperatures and lower luminosities -- not consistent with our CMFGEN models of the spectra during quiescence. This has implications for the star's luminosity and thus its evolutionary past, see Section \ref{discussion:classification}.

\subsection{Fundamental stellar wind parameters in quiescence state}
\label{results:physicalparameters_quiescence}

\begin{figure*}
\centering
\includegraphics[width=1\textwidth]{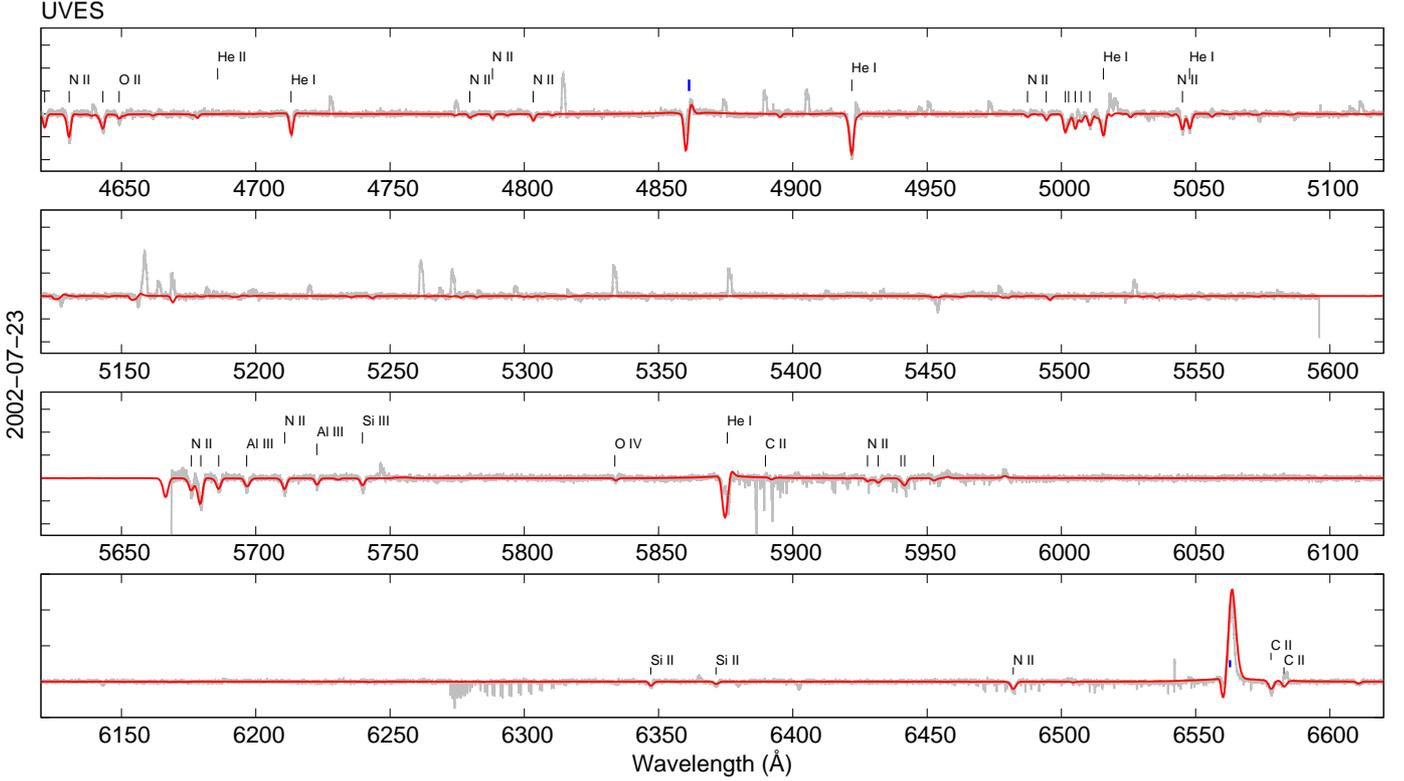}
     \caption{CMFGEN  best-fit model compared to the UVES spectrum obtained on 2002-07-23. The blue marks indicate hydrogen lines. The unfitted emission lines are nebular [\ion{Fe}{II}] lines.}\label{figure:CMFGEN_UVES}
\end{figure*}

\begin{table}
\caption{R71's fundamental stellar parameters during the star's quiescent phase (from our best-fit CMFGEN model). \label{table:parameters_quiescense}}
\begin{tabular}{lcc}
\hline\hline
Parameter &  value & error  \\ 
\hline
$T_\textnormal{\scriptsize{eff}}^a$	&   $15\,500$~K & $500$~K\\
$T_{*}^b$	&  16\,200~K	\\
log$(L_{*}/L_{\odot})$	&   $5.78$ &$0.02$ \\
$M_\textnormal{\scriptsize{bol}}$	&   $-9.65$ & $0.02$ \\
$\dot{M}$	 & $4.0 \times 10^{-6}~M_{\odot}$~yr$^{-1}$ & $0.5 \times 10^{-6}~M_{\odot}$~yr$^{-1}$ \\
log$(g/\textnormal{[cgs]})^c$ &	 $1.80$ & $0.1$ \\
$M$	& $27~M_{\odot}$ \\
$R_\textnormal{\scriptsize{phot}}$	&  107~$R_{\odot}$ \\
$R_{*}$	&  98~$R_{\odot}$ \\
$v_\mathrm{turb}$	&   $15$~km~s$^{-1}$ & $5$~km~s$^{-1}$ \\
$v_{\infty}$	&   190~km~s$^{-1}$ &  10~km~s$^{-1}$\\
$A_{V}$	&   0.6~mag & 0.1~mag\\
\hline
\multicolumn{3}{l}{$^a$ Parameters labelled ``$\textnormal{{eff}}$'' and ``$\textnormal{{phot}}$'' refer to $\tau_\textnormal{\scriptsize{ROSS}} = 2/3$.} \\
\multicolumn{3}{l}{$^b$ Parameters labelled ``$\textnormal{{*}}$'' refer to $\tau_\textnormal{\scriptsize{ROSS}} = 20$.} \\
\multicolumn{3}{l}{$^c$ Specified at $\tau_\textnormal{\scriptsize{ROSS}} = 2/3$.} \\
\end{tabular}
\end{table}

We used pre-outburst spectra obtained in 1984--2005 and CMFGEN atmosphere modeling to determine the classical stellar parameters of R71. The modeling criteria are described in Section \ref{obs:CMFGEN}. Small spectral variations, such as the strength of hydrogen lines, can be observed in this data set. However, for simplicity and lack of simultaneous photometry for each spectral epoch, we determine the average of the fundamental stellar parameters over the time span of the star's quiescent state. During the outburst, R71 has effective temperatures of $6000-6500$~K and is too cool to be modeled with the (hot star wind) code CMFGEN. 

Figure \ref{figure:CMFGEN_UVES} shows the 2002 UVES spectrum compared to our best-fit CMFGEN model. The other spectral epochs and their comparison with the model are in the Appendix \ref{appendix:models} (Figures \ref{figure:CMFGEN_CASPEC} to \ref{figure:CMFGEN_FEROS}).  Derived stellar parameters for R71's quiescent state are listed in Table \ref{table:parameters_quiescense}.
We find an effective temperature of $T_\textnormal{\scriptsize{eff}} = 15\,500 \pm 500$~K and a stellar luminosity of log$(L/L_{\odot}) = 5.78 \pm 0.02$ ($M_\textnormal{\scriptsize{bol}}  = -9.65$~mag).
We use the blue edge of the H$\alpha$ P\,Cyg absorption and other Balmer lines as a proxy for the wind terminal velocity. Since the absorption is not well-defined and filled in with extra emission this measurement gives only a lower limit of $v_\infty = 190$~km~s$^{-1}$. 
From the metallic resonance line \ion{Mg}{II} 2803, observed in ultraviolet IUE spectra obtained in 1981, 1985, and 1986, when the star was in a quiescent state, we determine a terminal velocity of $v_\infty = 203 \pm2$~km~s$^{-1}$. An additional higher velocity component is observed, which extends to $v_\infty = 281 \pm7$~km~s$^{-1}$.
The gravity is estimated to be log$(g/\textnormal{[cgs]}) = 1.80$ and the stellar mass to $27~M_{\odot}$. A mass-loss rate of $4.0 \times 10^{-6}~M_{\odot}$~yr$^{-1}$ for a clumping factor equal to 1 fits the hydrogen lines well. However, extended emission from surrounding material is apparent in the spectra and not all hydrogen and helium lines are matched equally well.

\subsection{Surface chemical abundances}
\label{results:abundances}

\begin{table}
\caption{Surface chemical abundances for R71.$^a$  Error estimates are on a 10--20\% level. For He the error may be as large as 50\%. \label{table:abundances}}
\begin{tabular}{cccc}
\hline\hline 
Species & Number fraction & Mass fraction$^b$ &  Z/Z$_{\textnormal{\scriptsize{LMC}}}$  \\ 
& (relative to H) & &   \\ 
\hline
H & 1.0 & 5.5e-01 &  0.745   \\
He & 0.2 & 4.4e-01 &  1.719   \\
C & 3.0e-05 & 2.0e-04 &  0.204   \\
N & 6.0e-04 & 4.6e-03 &  16.292   \\
O & 5.0e-05 & 4.4e-04 &  0.180  \\
\hline
\multicolumn{4}{l}{$^a$ Derived from the 1984--2005 pre-outburst spectra.} \\
\multicolumn{4}{l}{$^b$ LMC mass fraction: H = 7.38e-01, He = 2.56e-01, C =  9.79e-04,} \\
\multicolumn{4}{l}{\phantom{$^b$} N = 2.82e-04, O = 2.45e-03.} \\\end{tabular}
\end{table}

We determine abundances of H, He, and the CNO elements as described in Section \ref{obs:CMFGEN}. Table \ref{table:abundances} provides an overview of the derived number fraction relative to hydrogen, mass fraction, and the ratio to standard LMC abundances. R71's abundances imply a strong enrichment of the atmosphere with CNO processed material.

We find similar CNO abundances as \citet{1993SSRv...66..207L}. These authors determined the abundances using CASPEC data, which only provide a very limited wavelength range from 3740--5445. In addition to the CASPEC spectra, we use UVES and FEROS spectra, covering the entire wavelength region from 3280--9200\AA. Lennon et al.\ found number ratios of He/H = 0.43, O/N = 0.14, and C/N = 0.03, while we find number ratios of He/H = 0.20, O/N = 0.08, and C/N = 0.05. Determining the He abundance is difficult and the error bars are large, i.e., we can fit the spectra using number ratios of He/H between 0.10 and 0.3. We find that a microturbulence of $10-20$~km~s$^{-1}$ matches most lines, while the \ion{He}{I} 6678, 7065 lines are best matched with $v_{turb} \approx 30$~km~s$^{-1}$. These values correspond well to literature values of supergiants: \citet{2002A&A...395..519T} determined  microturbulences for M31 and Galactic supergiants in the range of 19~km~s$^{-1}$ to 35~km~s$^{-1}$ and \citet{2007A&A...471..625T} find microturbulences in Galactic, LMC, and SMC B-type stars between 0~km~s$^{-1}$ to 20~km~s$^{-1}$.

\section{Discussion}
\label{discussion}

\subsection{The evolutionary state}
\label{discussion:classification}

\begin{figure*}
\centering
\resizebox{\hsize}{!}{\includegraphics{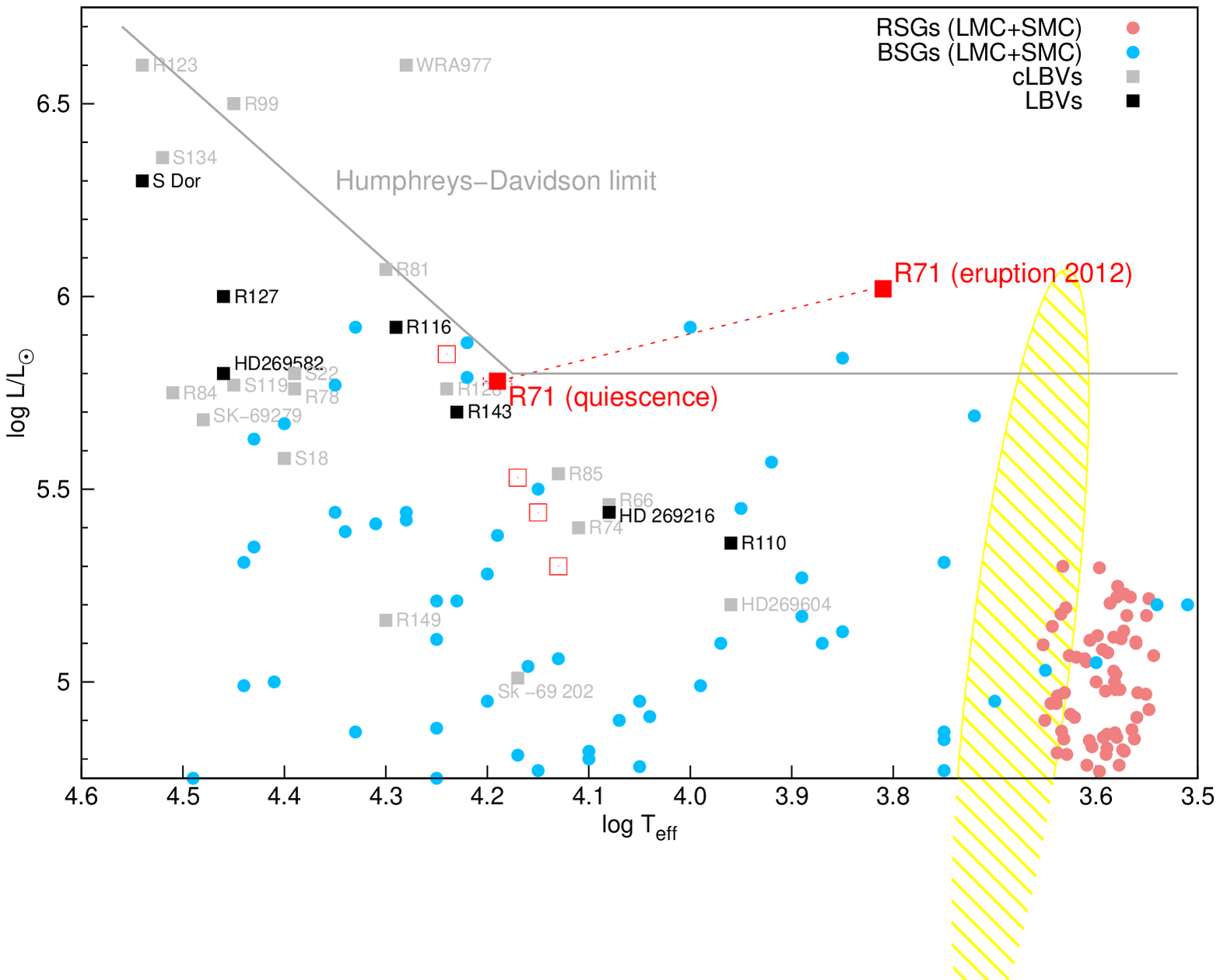}}
     \caption{Schematic upper HR diagram of the LMC. The solid gray curve is the upper luminosity boundary for the LMC \citep{1979ApJ...232..409H}. Confirmed LBVs (black) and candidate LBVs (gray) are displayed at their minimum phases \citep{1986A&A...163..119Z,1996A&A...315..510Z,2001A&A...366..508V,2002A&A...389..931T,2004ApJ...610.1021E,2014A&A...565A..27H}. Blue and Red Supergiants are indicated with blue and red filled circles \citep{2004A&A...417..217T,2005A&A...434..677T,2006ApJ...645.1102L,2008A&A...479..541H}.
We show the locations of R71 during its quiescent state and its current outburst (filled red squares, parameters from this paper and \citealt{2013A&A...555A.116M}). If R71 really increased its bolometric luminosity during the outburst as indicated in this figure requires more detailed modeling beyond the scope of this paper. Previous literature values of R71's quiescent state are also shown (open red squares). The dashed yellow ellipse indicates the extension of the Cepheid instability strip.} \label{figure:HRdiagram}
\end{figure*}

To better understand the LBV instability phenomenon it is important to interpret the evolutionary state of LBVs and their evolutionary history. In the case of R71, it is crucial to know the luminosity, because it may imply different evolutionary paths (see Section \ref{intro}).
However, there is ambiguity in the literature regarding the luminosity of R71 \citep{1981A&A...103...94W,1993SSRv...66..207L}. The reason for this uncertainty lies in the different extinction values and effective temperatures (and thus bolometric corrections) assumed by different authors (Table 2 in \citealt{2013A&A...555A.116M}). 
Support for the classification of R71 as a less luminous LBV came from the fact that its visual light and temperature during the 1970s outburst had much smaller variations than is observed for the more luminous LBVs  \citep{1974MNRAS.168..221T,1981A&A...103...94W}. However, the current outburst displays all signs of an outburst of a classical LBV and it appears not to be uncommon that outbursts of the same star can have different magnitudes \citep{2017AJ....154...15W}. The large differences in S\,Doradus outbursts for the same star should be taken into account when discussing potential causes for the instability. It also renders the classification scheme based on outburst magnitude futile.

Figure \ref{figure:HRdiagram} shows the location of R71 in the HR diagram during quiescence (determined in this paper) and during its current outburst state (determined in \citealt{2013A&A...555A.116M}).  The four open red squares indicate literature parameters \citep{1981A&A...103...94W,1982A&A...112...61V,1988A&AS...74..453V,1993SSRv...66..207L}. Their substantial spread across the HR diagram demonstrates the large errors associated with stellar parameters published for LBVs -- even when the distance is well known as in the case of R71.
With our CMFGEN atmosphere modeling analysis we can finally settle the luminosity debate. R71 occupies the region of classical LBVs in the HR diagram, but at the lower luminosity end. A previous RSG phase, which in particular composition studies support \citep{1999A&A...341L..67V,2010AJ....139...68V,2010A&A...518L.142B}, cannot be ruled out.

R71 has a strong surface enrichment in CNO processed material. It is, however, not possible to constrain its evolutionary state or deduce its evolutionary history based on it. Different (not well constrained) factors such as mass loss and mixing affect the  evolution of the surface abundances strongly \citep{2013LNP...865....3M} and the correct treatment of convection in stellar evolutionary codes is still uncertain \citep{2014MNRAS.439L...6G}. Georgy et al.\ show that the surface chemical abundances of Blue Supergiants are very sensitive to the way convection is modeled in the stellar evolution code (Ledoux criterion versus Schwarzschild criterion).

\citet{2015MNRAS.447..598S} suggest that LBVs are the rejuvenated mass gainer products of massive star binary evolution, while \citet{2016ApJ...825...64H} argue that LBVs are predominantly the products of single star evolution. 
R71 is the most isolated LBV in the LMC from nearby O-type stars \citep{2015MNRAS.447..598S} and \citet{2016arXiv161105504L} compute its space velocity using the Gaia TGAS proper motion catalog. Both studies support the case that R71 cannot have evolved as a single star, because it lies too far from massive (OB-type) star forming complexes to have arrived at its current position during a lifetime as a single star. In the case of a binary, mass transfer from this putative companion may have rejuvenated R71. Maybe its S\,Doradus outbursts and the oscillatory variability discussed in Section \ref{results:cycle} could be triggered or modulated by this putative close companion.

\subsection{Instabilities in the atmosphere during the S\,Doradus outburst}
\label{discussion:pulsations}

During quiescent state, R71 shows microvariations of $\Delta m_{\textnormal{\scriptsize{V}}}\sim0.1$~mag on timescales of 14--100~days attributed to large turbulent elements and dynamical instabilities in the photosphere or to pulsations \citep{1985A&A...153..163V,1988A&AS...74..453V,1997A&AS..124..517V,1998A&A...335..605L}.
In this paper, we report on larger amplitude and longer period oscillations in R71's light curve and spectrum during the current outburst. 
The spectral variations, i.e., the occurrence of double absorption lines, have a cycle length of about 850~d. The visual light curve displays sinusoidal variations of $\sim0.2$~mag with a period of about 425~d, hinting at period doubling. Variations in R71's light curve are also seen during the rise in brightness in 2007 and 2008, but with a shorter timescale. This is qualitatively in agreement with pulsation because the star's mean density was higher then.

R71 falls in the LBV instability strip in the HR diagram during its quiescent phase, but does not cross any other known location of pulsating variables during its outburst phase, see Figure 1 in \citet{2016MNRAS.458.1352J}. However, the Cepheid instability strip may extend to higher luminosities beyond the RV\,Tau variables (meshed yellow ellipse in Figure \ref{figure:HRdiagram}). 
The much expanded, cool envelope during the outburst now engulfs the \ion{Fe}{II}/\ion{Fe}{III} ionization zone, which could provide driving through the $\kappa$-mechanism.

RV\,Tau variables are pulsating variables \citep{2007uvs..book.....P}. They are post-Asymptotic Giant Branch stars with masses close to solar mass and spectral types ranging from early F to early K. The reasons for their variability are still not entirely clear, but are supposedly seated in the outer photospheric layers. They may be caused by multiple pulsation modes present in the star or nonlinear or chaotic behavior. 
Indeed, R71's photometric and spectroscopic variations during outburst are similar to what is observed in RV\,Tau variables. The light curves of RV\,Tau stars are modulated in an irregular way with period doubling due to a resonance between the fundamental mode and the first overtone (e.g., \citealt{1983A&A...117..352T,1990ApJ...355..590M,1996MNRAS.279..949P}), which may also be the case for R71. Also, double absorption lines have been observed in RV\,Tau stars, attributed to shock waves within the atmosphere \citep{1989A&A...215..316G}.
Pulsations at such high stellar luminosities acting on a small amount of mass in the outer layers, are expected to be non-linear, which accounts for the irregular behavior of the visual light curve.

While RV\,Tau variables cross the Cepheid instability
during their evolution \citep{2002PASP..114..689W}, S\,Doradus outbursts do not represent evolutionary transitions in the HR diagram. During outbursts, LBVs only appear to be relocated in the HR diagram, but their true evolutionary status does not change. The inner structure of an LBV in outburst might thus be far away from an evolutionary equilibrium and any relation to either Cepheid or RV\,Tau stars can only be based on phenomenology.  It is also not clear if the phenomena observed for evolved lower mass stars can be extended to the atmospheres of LBVs during S\,Doradus outbursts. However, similar to RV\,Tau variables, LBVs in outburst have very extended, cool, low-mass outer atmospheres. 

\citet{1980A&A....90..311M} discusses supergiant pulsations with respect to their amplitude and period distribution across the HR diagram and their potential driving mechanisms. The author shows the importance of external convective zones present in these stars and discusses possibilities for non-radial oscillations resulting from either $g^-$ modes in the presence of an extra-restoring force like rotation or the excitation of $g^+$ modes. As stated by \citet{1985A&A...153..163V}, R71's microvariations during quiescent state fit well into the period-luminosity relation (equation 7 and Figure 6 in \citealt{1980A&A....90..311M}; a period of about 60~d is expected). Also, the observed oscillations during the current outburst with much larger amplitude and timescale fulfill well the period-luminosity relation; a period of 780~d is expected compared to the observed 850~d. This suggests oscillations in the convective outer zones of R71. We note that other LBVs in outburst show similar oscillations in their light curve, see the case of R127 (Figure 1 in \citealt{2017AJ....154...15W}). How the extended atmosphere of LBVs act in terms of convection and convective instabilities requires detailed theoretical attention.  Under the conditions of extended LBV atmospheres, the classical $\kappa$ mechanism could be very efficient, i.e., minor changes in the temperature could cause large changes in opacity.

\section{Conclusion}
\label{conclusion}

R71 was well-observed over the last decades covering its quiescent phases, its 1970--1977 S\,Doradus outburst, and its current outburst. The available multi-wavelength data sets give us a unique opportunity to better understand the LBV phenomenon. R71's current S\,Doradus outburst has a much larger amplitude and longer duration than its previous outbursts in the 1970s and the first half of the 20th century. Consecutive outbursts, within decades, of the same star can be of rather different magnitude and a classification scheme of LBVs based on one outburst is thus insufficient. 

We determined R71's fundamental stellar parameters using pre-outburst spectra and atmosphere models computed with CMFGEN. The star's effective temperature is $T_\textnormal{\scriptsize{eff}} = 15\,500 \pm 500$~K and its stellar luminosity is log$(L/L_{\odot}) = 5.78 \pm 0.02$. R71 is thus at the lower luminosity end of the classical LBVs during quiescence. Other stellar wind parameters and abundances are provided in the paper. We also present the entire SED from the ultra-violet to the mid-infrared of the star in outburst. Mid-infrared observations suggest an increase of hot dust during the current outbursts and cooling of dust and grain evolution in the most outer regions.

R71's visual light curve and its spectrum show oscillations during the current S\,Doradus outburst, resembling phenomena observed in the lower-mass RV\,Tau stars. The occurrence of double absorption lines may result from shocks in the atmosphere and the variations in the light curve may hint at period doubling.
The timescale of these variations fit well into the period-luminosity relation for supergiants \citep{1980A&A....90..311M}, which suggests oscillations in the convective outer zones of R71.
Further analysis and theoretical modeling is required to investigate these oscillations and also to constrain the mechanisms of the LBV phenomenon and the evolutionary state of LBVs.

\begin{acknowledgements} We acknowledge with thanks the variable star observations from the AAVSO International Database contributed by observers worldwide and used in this research \url{https://www.aavso.org}. ASAS data are available at \url{http://www.astrouw.edu.pl/asas/}.
The IUE data used in this paper were obtained from the Mikulski Archive for Space Telescopes (MAST). STScI is operated by the Association of Universities for Research in Astronomy, Inc., under NASA contract NAS5-26555. Support for MAST for non-HST data is provided by the NASA Office of Space Science via grant NNX09AF08G and by other grants and contracts. This publication makes use of data products from the Wide-field Infrared Survey Explorer, which is a joint project of the University of California, Los Angeles, and the Jet Propulsion Laboratory/California Institute of Technology, funded by the National Aeronautics and Space Administration. This research has made use of the NASA/IPAC Infrared Science Archive, which is operated by the Jet Propulsion Laboratory, California Institute of Technology, under contract with the National Aeronautics and Space Administration. This research has made use of NASA's Astrophysics Data System. CMFGEN can be downloaded at \url{http://kookaburra.phyast.pitt.edu/hillier/web/CMFGEN.htm}.  This research made use of Astropy, a community-developed core Python package
  for Astronomy \citep{2013A&A...558A..33A}. CA acknowledges support from FONDECYT grant No.\ 3150463 and from the Ministry of
Economy, Development, and Tourism's Millennium Science Initiative through grant IC120009, awarded to The Millennium Institute
of Astrophysics, MAS.
\end{acknowledgements}

\bibliographystyle{aa}
\bibliography{ms}

\begin{appendix}

\section{CMFGEN models}
\label{appendix:models}

\begin{figure*}
\centering
\includegraphics[width=0.95\textwidth]{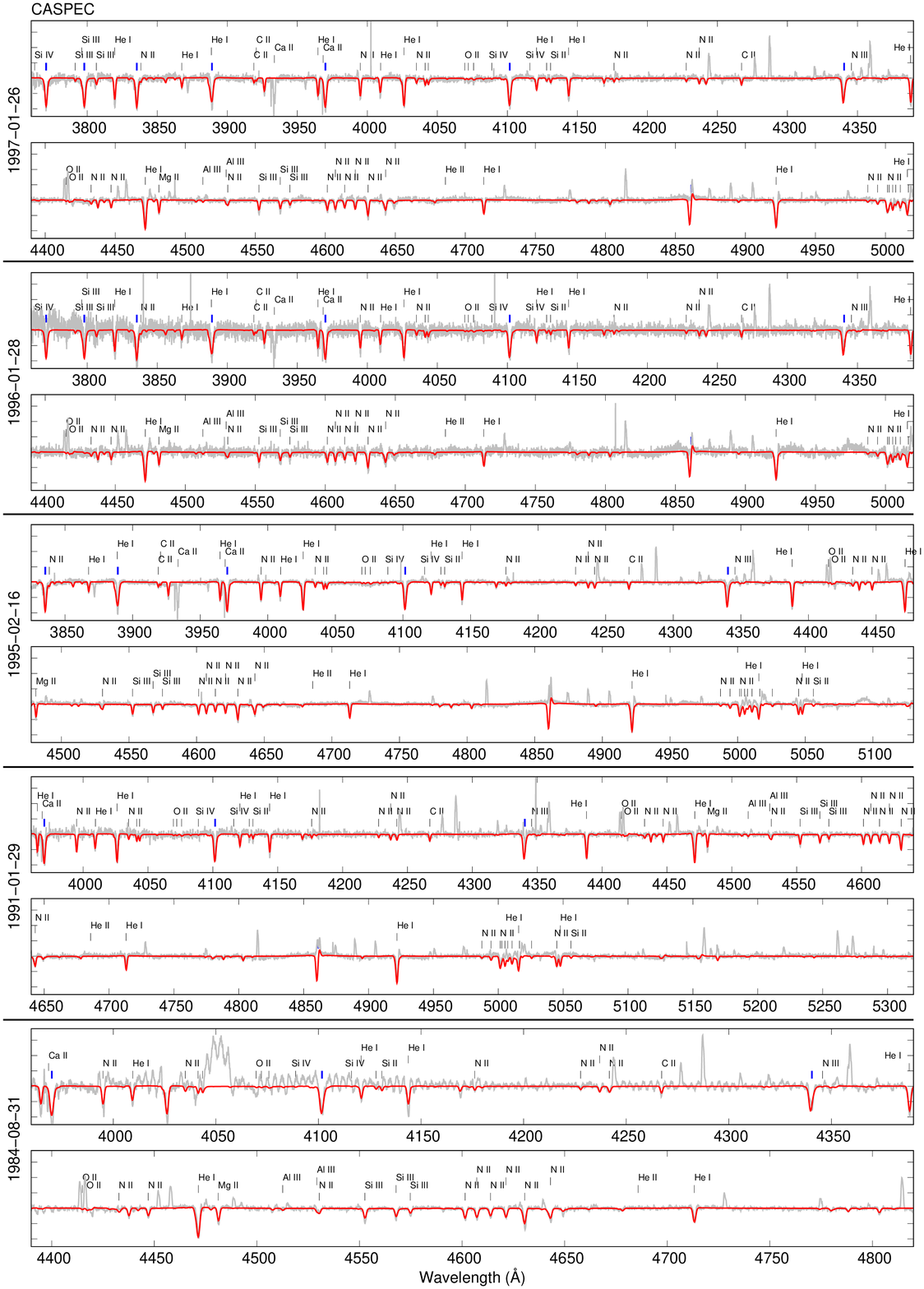}
     \caption{CMFGEN best-fit model compared to the CASPEC data from 1984 to 1997. The many emission lines are nebular [\ion{Fe}{II}] and [\ion{N}{II}] lines. The blue marks indicate hydrogen lines.}\label{figure:CMFGEN_CASPEC}
\end{figure*}

\begin{figure*}
\centering
\includegraphics[width=0.95\textwidth]{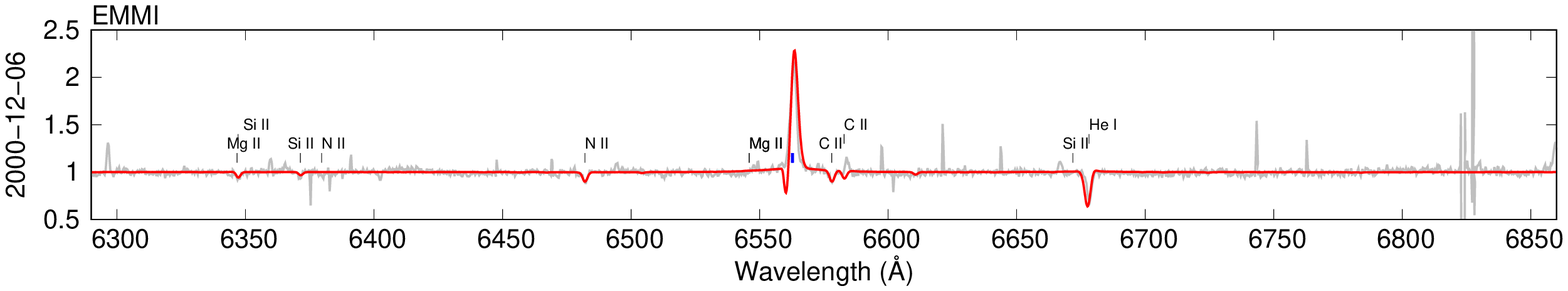}
     \caption{CMFGEN best-fit model compared to the EMMI data in 2000.}\label{figure:CMFGEN_EMMI}
\end{figure*}

\begin{figure*}
\centering
\includegraphics[width=0.95\textwidth]{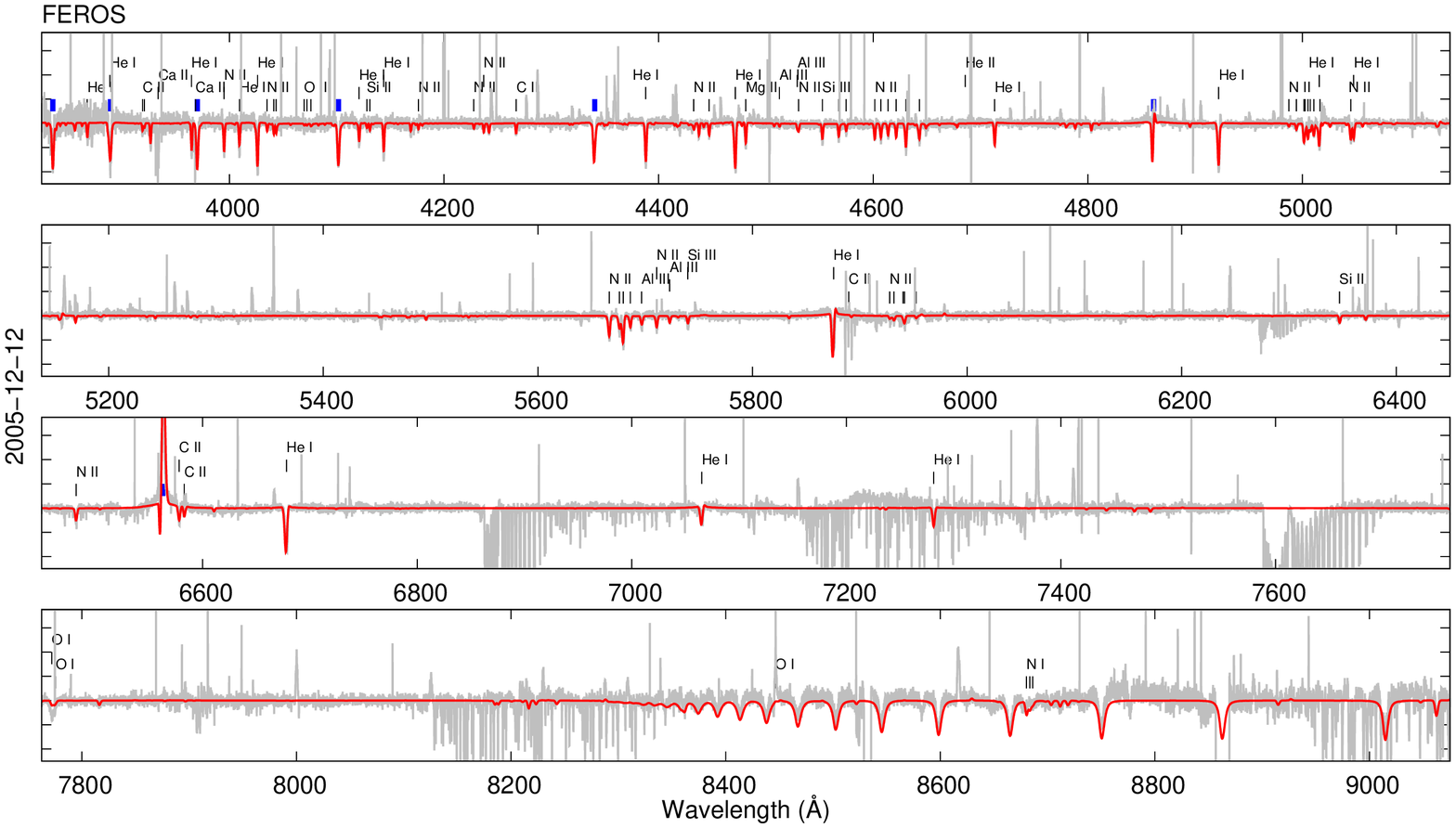}
     \caption{CMFGEN atmosphere models compared to the FEROS spectra obtained on 2005-12-12.}\label{figure:CMFGEN_FEROS}
\end{figure*}

\section{Journal of spectroscopic observations}
\label{appendix:journal}

\begin{longtable}{cccccc}
\caption{Journal of spectroscopic observations.\label{table:journal}}\\
\hline\hline \setlength{\tabcolsep}{4pt}
Date & MJD  & Instrument   & Wavelength Range & \centering{Spectral Resolving} & Total Exposure Time   \\ 
 & (days)  & & (\AA) &  \multicolumn{1}{c}{Power} & (s)  \\ 
\hline
\endfirsthead
\caption{continued.}\\
\hline\hline
Date & MJD  & Instrument   & Wavelength Range & \centering{Spectral Resolving} & Total Exposure Time   \\ 
 & (days)  & & (\AA) &  \multicolumn{1}{c}{Power} & (s)  \\ 
 \hline
\endhead
\hline
\endfoot

1984-08-31 & 45943 & CASPEC & 3895--4970 & 20\,000 &  1500  \\
1991-01-29 & 48285 & CASPEC & 3920--5445 & 20\,000 &  1200  \\
1995-02-16 & 49764 & CASPEC & 3815--5140 & 20\,000 &  3000  \\
1996-01-28 & 50110 & CASPEC & 3740--5045 & 20\,000 &  756  \\
1997-01-26 & 50474 & CASPEC & 3765--5045 & 20\,000 &  1200  \\
\hline
2000-12-06 & 51884.2 & EMMI & 6240--6870 & 20\,000 &  600,1800  \\
\hline
2002-07-23	&	52478.4	&	UVES\tablefootmark{1}    	&	3\,280--4\,560 	& 40\,000 &	 	360 \\
2002-07-23	&	52478.4	&	UVES    	&	4\,580--6\,690	& 40\,000 &	3x110 \\
2002-08-03	&	52489.4	&	UVES    	&	3\,280--4\,560  & 40\,000 &	360 \\
2002-08-03	&	52489.4	&	UVES    	&	4\,580--6\,690	& 40\,000 &	3x110 \\
2002-08-12	&	52498.4	&	UVES    	&	3\,280--4\,560	& 40\,000 &	360 \\
2002-08-12	&	52498.4	&	UVES    	&	4\,580--6\,690	& 40\,000 &	3x110 \\
2002-09-04    &	52521.4	&	UVES    	&	3\,280--4\,560	& 40\,000 &	360 \\
2002-09-04	&	52521.4	&	UVES    	&	4\,580--6\,690	& 40\,000 &	3x110 \\
2002-09-10	&	52527.4	&	UVES    	&	3\,280--4\,560	& 40\,000 &	360 \\
2002-09-10	&	52527.4	&	UVES    	&	4\,580--6\,690	& 40\,000 &	3x110 \\
2002-09-18	&	52535.3	&	UVES    	&	3\,280--4\,560	& 40\,000 &	360 \\
2002-09-18    &	52535.3	&	UVES    	&	4\,580--6\,690	& 40\,000 &	3x110 \\
2002-09-22	&	52539.3	&	UVES    	&  3\,280--4\,560	& 40\,000 &	360 \\
2002-09-22	&	52539.3	&	UVES    	&	4\,580--6\,690	& 40\,000 &	3x110 \\
2002-09-30	&	52547.4	&	UVES    	&	3\,280--4\,560	& 40\,000 &	360 \\
2002-09-30	&	52547.4	&	UVES    	&	4\,580--6\,690	& 40\,000 &	3x110 \\
2002-10-06	&	52553.4	&	UVES    	&	3\,280--4\,560	& 40\,000 &	360 \\
2002-10-06	&	52553.4	&	UVES    	&	4\,580--6\,690	& 40\,000 &	3x110 \\
\hline
2005-12-12	&	53716.2	&	FEROS   	&	3\,500--9\,200	& 48\,000 &	2x450	\\
2007-02-22	&	54153.2	&	FEROS   	&	3\,500--9\,200	& 48\,000 &	3x500	\\
2011-02-13	& 55605.0	&	FEROS   	&	3\,500--9\,200	& 48\,000 &	900	\\
2011-03-20	& 55641.0	&	FEROS   	&	3\,500--9\,200	& 48\,000 &	1800	\\
2011-05-15	&	55697.0	&	FEROS   	&	3\,500--9\,200	& 48\,000 &	1800	\\
2012-06-19	&	56097.4	&	FEROS   	&	3\,500--9\,200	& 48\,000 &	1800	\\
2014-11-29	&	56990.1	&	FEROS   	&	3\,500--9\,200	& 48\,000 &	2x960	\\
2015-10-16	&	57311.3	&	FEROS   	&	3\,500--9\,200	& 48\,000 &	2x500	\\
\hline
2012-08-26	&	56165.3	&	X-shooter\tablefootmark{2}	&	10\,240--24\,800	& 10\,000 &	3x60, 2x10	\\
2012-08-26 	&	56165.3	&	X-shooter	&	3\,000--5\,595	& 9\,000 &	3x120, 2x20	\\
2012-08-26 	&	56165.3	&	X-shooter	&	5\,595--10\,240	& 17\,000 &	3x120, 2x20	\\

2012-09-04	&	56174.3	&	X-shooter	&	10\,240--24\,800	& 10\,000 &	3x60, 2x10	\\
2012-09-04 	&	56174.3	&	X-shooter	&	3\,000--5\,595	& 9\,000 &	3x120, 2x20	\\
2012-09-04 	&	56174.3	&	X-shooter	&	5\,595--10\,240	& 17\,000 &	3x120, 2x20	\\

2012-11-01	&	56232.2	&	X-shooter	&	10\,240--24\,800	& 10\,000 &	3x60, 2x10	\\
2012-11-01 	&	56232.2	&	X-shooter	&	3\,000--5\,595	& 9\,000 &	3x120, 2x20	\\
2012-11-01 	&	56232.2	&	X-shooter	&	5\,595--10\,240	& 17\,000 &	3x120, 2x20	\\

2013-02-02	&	56325.0	&	X-shooter	&	10\,240--24\,800	& 10\,000 &	3x60, 2x10	\\
2013-02-02 	&	56325.0	&	X-shooter	&	3\,000--5\,595	& 9\,000 &	3x120, 2x20	\\
2013-02-02 	&	56325.0	&	X-shooter	&	5\,595--10\,240	& 17\,000 &	3x120, 2x20	\\

2013-03-01	&	56352.1	&	X-shooter	&	10\,240--24\,800	& 10\,000 &	3x60, 2x10	\\
2013-03-01 	&	56352.1	&	X-shooter	&	3\,000--5\,595	& 9\,000 &	3x120, 2x20	\\
2013-03-01 	&	56352.1	&	X-shooter	&	5\,595--10\,240	& 17\,000 &	3x120, 2x20	\\

2013-08-04	&	56508.4	&	X-shooter	&	10\,240--24\,800	& 10\,000 &	4x60, 2x10	\\
2013-08-04 	&	56508.4	&	X-shooter	&	3\,000--5\,595	& 9\,000 &	4x120, 2x20	\\
2013-08-04 	&	56508.4	&	X-shooter	&	5\,595--10\,240	& 17\,000 &	4x120, 2x20	\\

2013-09-19	&	56554.3	&	X-shooter	&	10\,240--24\,800	& 10\,000 &	4x60, 2x10	\\
2013-09-19 	&	56554.3	&	X-shooter	&	3\,000--5\,595	& 9\,000 &	4x120, 2x20	\\
2013-09-19 	&	56554.3	&	X-shooter	&	5\,595--10\,240	& 17\,000 &	4x120, 2x20	\\

2013-10-24	&	56589.3	&	X-shooter	&	10\,240--24\,800	& 10\,000 &	3x60, 2x10	\\
2013-10-24 	&	56589.3	&	X-shooter	&	3\,000--5\,595	& 9\,000 &	3x120, 2x20	\\
2013-10-24 	&	56589.3	&	X-shooter	&	5\,595--10\,240	& 17\,000 &	3x120, 2x20	\\

2013-11-12	&	56608.1	&	X-shooter	&	10\,240--24\,800	& 10\,000 &	3x60, 2x10	\\
2013-11-12 	&	56608.1	&	X-shooter	&	3\,000--5\,595	& 9\,000 &	3x120, 2x20	\\
2013-11-12 	&	56608.1	&	X-shooter	&	5\,595--10\,240	& 17\,000 &	3x120, 2x20	\\

2013-12-22	&	56648.3	&	X-shooter	&	10\,240--24\,800	& 10\,000 &	3x60, 2x10	\\
2013-12-22 	&	56648.3	&	X-shooter	&	3\,000--5\,595	& 9\,000 &	3x120, 2x20	\\
2013-12-22 	&	56648.3	&	X-shooter	&	5\,595--10\,240	& 17\,000 &	3x120, 2x20	\\

2014-02-01	&	56689.2	&	X-shooter	&	10\,240--24\,800	& 10\,000 &	3x60, 2x10	\\
2014-02-01 	&	56689.2	&	X-shooter	&	3\,000--5\,595	& 9\,000 &	3x120, 2x20	\\
2014-02-01 	&	56689.2	&	X-shooter	&	5\,595--10\,240	& 17\,000 &	3x120, 2x20	\\

2014-11-16	&	56977.1	&	X-shooter	&	10\,240--24\,800	& 10\,000 &	3x60, 2x10	\\
2014-11-16 	&	56977.1	&	X-shooter	&	3\,000--5\,595	& 9\,000 &	3x120, 2x20	\\
2014-11-16 	&	56977.1	&	X-shooter	&	5\,595--10\,240	& 17\,000 &	3x120, 2x20	\\

2014-12-27	&	57018.0	&	X-shooter	&	10\,240--24\,800	& 10\,000 &	3x60, 2x10	\\
2014-12-27 	&	57018.0	&	X-shooter	&	3\,000--5\,595	& 9\,000 &	3x120, 2x20	\\
2014-12-27 	&	57018.0	&	X-shooter	&	5\,595--10\,240	& 17\,000 &	3x120, 2x20	\\

2015-03-02	&	57083.1	&	X-shooter	&	10\,240--24\,800	& 10\,000 &	3x60, 2x10	\\
2015-03-02 	&	57083.1	&	X-shooter	&	3\,000--5\,595	& 9\,000 &	3x120, 2x20	\\
2015-03-02 	&	57083.1	&	X-shooter	&	5\,595--10\,240	& 17\,000 &	3x120, 2x20	\\

2015-10-15	&	57310.3	&	X-shooter	&	10\,240--24\,800	& 10\,000 &	3x60, 2x10	\\
2015-10-15 	&	57310.3	&	X-shooter	&	3\,000--5\,595	& 9\,000 &	3x120, 2x20	\\
2015-10-15 	&	57310.3	&	X-shooter	&	5\,595--10\,240	& 17\,000 &	3x120, 2x20	\\

2015-12-18	&	57374.3	&	X-shooter	&	10\,240--24\,800	& 10\,000 &	3x60, 2x10	\\
2015-12-18 	&	57374.3	&	X-shooter	&	3\,000--5\,595	& 9\,000 &	3x120, 2x20	\\
2015-12-18 	&	57374.3	&	X-shooter	&	5\,595--10\,240	& 17\,000 &	3x120, 2x20	\\

2016-02-19	&	57437.1	&	X-shooter	&	10\,240--24\,800	& 10\,000 &	3x60, 2x10	\\
2016-02-19 	&	57437.1	&	X-shooter	&	3\,000--5\,595	& 9\,000 &	3x120, 2x20	\\
2016-02-19 	&	57437.1	&	X-shooter	&	5\,595--10\,240	& 17\,000 &	3x120, 2x20	\\

2016-04-06	&	57484.0	&	X-shooter	&	10\,240--24\,800	& 10\,000 &	4x60, 2x10	\\
2016-04-06 	&	57484.0	&	X-shooter	&	3\,000--5\,595	& 9\,000 &	4x120, 2x20	\\
2016-04-06 	&	57484.0	&	X-shooter	&	5\,595--10\,240	& 17\,000 &	4x120, 2x20	\\

2016-08-03	&	57603.4	&	X-shooter	&	10\,240--24\,800	& 10\,000 &	4x60, 2x10	\\
2016-08-03 	&	57603.4	&	X-shooter	&	3\,000--5\,595	& 9\,000 &	4x120, 2x20	\\
2016-08-03 	&	57603.4	&	X-shooter	&	5\,595--10\,240	& 17\,000 &	4x120, 2x20	\\

2016-10-06	&	57667.4	&	X-shooter	&	10\,240--24\,800	& 10\,000 &	4x60, 2x10	\\
2016-10-06 	&	57667.4	&	X-shooter	&	3\,000--5\,595	& 9\,000 &	4x120, 2x20	\\
2016-10-06 	&	57667.4	&	X-shooter	&	5\,595--10\,240	& 17\,000 &	4x120, 2x20	\\

2017-01-02	&	57755.3	&	X-shooter	&	10\,240--24\,800	& 10\,000 &	4x60, 2x10	\\
2017-01-02 	&	57755.3	&	X-shooter	&	3\,000--5\,595	& 9\,000 &	4x120, 2x20	\\
2017-01-02 	&	57755.3	&	X-shooter	&	5\,595--10\,240	& 17\,000 &	4x120, 2x20	\\

2017-04-04	&	57847.0	&	X-shooter	&	10\,240--24\,800	& 10\,000 &	4x60, 2x10	\\
2017-04-04 	&	57847.0	&	X-shooter	&	3\,000--5\,595	& 9\,000 &	4x120, 2x20	\\
2017-04-04 	&	57847.0	&	X-shooter	&	5\,595--10\,240	& 17\,000 &	4x120, 2x20	\\

2017-08-03	&	57968.3	&	X-shooter	&	10\,240--24\,800	& 10\,000 &	8x60, 4x10	\\
2017-08-03 	&	57968.3	&	X-shooter	&	3\,000--5\,595	& 9\,000 &	8x120, 4x20	\\
2017-08-03 	&	57968.3	&	X-shooter	&	5\,595--10\,240	& 17\,000 &	8x120, 4x20	\\

\hline 
\multicolumn{6}{l}{$^1$ UVES slit width: 1{\arcsec}.} \\
\multicolumn{6}{l}{$^2$ X-shooter slit width: 0\farcs5/0\farcs4/0\farcs4 for the UVB/VIS/NIR arms and long exposures, 5{\arcsec}/5{\arcsec}/5{\arcsec} for the short exposures.} \\
\end{longtable}

\end{appendix}

\end{document}